\documentstyle[12pt,a4]{article}

\begin{document}

\newcommand{\nc}[2]{\newcommand{#1}{#2}}
\newcommand{\ncx}[3]{\newcommand{#1}[#2]{#3}}
\ncx{\pr}{1}{#1^{\prime}}
\nc{\nl}{\newline}
\nc{\np}{\newpage}
\nc{\nit}{\noindent}
\nc{\be}{\begin{equation}}
\nc{\ee}{\end{equation}}
\nc{\ba}{\begin{array}}
\nc{\ea}{\end{array}}
\nc{\bea}{\begin{eqnarray}}
\nc{\eea}{\end{eqnarray}}
\nc{\nb}{\nonumber}
\nc{\dsp}{\displaystyle}
\nc{\bit}{\bibitem}
\nc{\ct}{\cite}
\ncx{\dd}{2}{\frac{\partial #1}{\partial #2}}
\nc{\pl}{\partial}
\nc{\dg}{\dagger}
\nc{\cH}{{\cal H}}
\nc{\cL}{{\cal L}}
\nc{\cD}{{\cal D}}
\nc{\cF}{{\cal F}}
\nc{\cG}{{\cal G}}
\nc{\cJ}{{\cal J}}
\nc{\cQ}{{\cal Q}}
\nc{\tB}{\tilde{B}}
\nc{\tD}{\tilde{D}}
\nc{\tH}{\tilde{H}}
\nc{\tR}{\tilde{R}}
\nc{\tZ}{\tilde{Z}}
\nc{\tg}{\tilde{g}}
\nc{\tog}{\tilde{\og}}
\nc{\tGam}{\tilde{\Gam}}
\nc{\tPi}{\tilde{\Pi}}
\nc{\tcD}{\tilde{\cD}}
\nc{\tcQ}{\tilde{\cQ}}
\nc{\ag}{\alpha}
\nc{\bg}{\beta}
\nc{\gam}{\gamma}
\nc{\Gam}{\Gamma}
\nc{\bgm}{\bar{\gam}}
\nc{\del}{\delta}
\nc{\Del}{\Delta}
\nc{\eps}{\epsilon}
\nc{\ve}{\varepsilon}
\nc{\zg}{\zeta}
\nc{\th}{\theta}
\nc{\vt}{\vartheta}
\nc{\Th}{\Theta}
\nc{\kg}{\kappa}
\nc{\lb}{\lambda}
\nc{\Lb}{\Lambda}
\nc{\ps}{\psi}
\nc{\Ps}{\Psi}
\nc{\sg}{\sigma}
\nc{\spr}{\pr{\sg}}
\nc{\Sg}{\Sigma}
\nc{\rg}{\rho}
\nc{\fg}{\phi}
\nc{\Fg}{\Phi}
\nc{\vf}{\varphi}
\nc{\og}{\omega}
\nc{\Og}{\Omega}
\nc{\dL}{\del_{\Lambda}}
\nc{\Kq}{\mbox{$K(\vec{q},t|\pr{\vec{q}\,},\pr{t})$ }}
\nc{\Kp}{\mbox{$K(\vec{q},t|\pr{\vec{p}\,},\pr{t})$ }}
\nc{\vq}{\mbox{$\vec{q}$}}
\nc{\qp}{\mbox{$\pr{\vec{q}\,}$}}
\nc{\vp}{\mbox{$\vec{p}$}}
\nc{\va}{\mbox{$\vec{a}$}}
\nc{\vb}{\mbox{$\vec{b}$}}
\nc{\Ztwo}{\mbox{\sf Z}_{2}}
\nc{\Tr}{\mbox{Tr}}
\nc{\lh}{\left(}
\nc{\rh}{\right)}
\nc{\ld}{\left.}
\nc{\rd}{\right.}
\nc{\nil}{\emptyset}
\nc{\bor}{\overline}
\nc{\ha}{\hat{a}}
\nc{\da}{\hat{a}^{\dg}}
\nc{\hb}{\hat{b}}
\nc{\db}{\hat{b}^{\dg}}
\nc{\hc}{\hat{c}}
\nc{\he}{\hat{e}}
\nc{\hp}{\hat{p}}
\nc{\hx}{\hat{x}}
\nc{\hH}{\hat{H}}
\nc{\hI}{\hat{\mbox{I}}}
\nc{\hK}{\hat{K}}
\nc{\hM}{\hat{M}}
\nc{\hN}{\hat{N}}
\nc{\hQ}{\hat{Q}}
\nc{\hag}{\hat{\ag}}
\nc{\hbg}{\hat{\bg}}
\nc{\hgm}{\hat{\gam}}
\nc{\hch}{\hat{\chi}}
\nc{\hps}{\hat{\psi}}
\nc{\hDel}{\hat{\Del}}
\nc{\hOg}{\hat{\Og}}
\nc{\pz}{z^{\prime}}
\nc{\pZ}{Z^{\prime}}
\ncx{\abs}{1}{\left| #1 \right|}
\nc{\vs}{\vspace{2ex}}

\pagestyle{empty}

\begin{flushright}
\begin{tabular}{c}
NIKHEF-H/95-050 \\
hep-th/9508136
\end{tabular}
\end{flushright}
\vspace{5ex}

\begin{center}
{\LARGE {\bf Propagators and Path Integrals }} \\

\vspace{5ex}

{\large J.W.\ van Holten} \\
        NIKHEF-H, P.O.\ Box 41882 \\
        1009 DB Amsterdam NL \\

August 22, 1995 \\

\vspace{20ex}

{\bf Abstract} \\
\end{center}

\nit
{\small
Path-integral expressions for one-particle propagators in scalar and fermionic
field theories are derived, for arbitrary mass. This establishes a direct
connection between field theory and specific classical point-particle models.
The role of world-line reparametrization invariance of the classical action
and the implementation of the corresponding BRST-symmetry in the quantum
theory are discussed. The presence of classical world-line supersymmetry is
shown to lead to an unwanted doubling of states for massive spin-1/2
particles. The origin of this phenomenon is traced to a `hidden' topological
fermionic excitation. A different formulation of the pseudo-classical mechanics
using a bosonic representation of $\gam_5$ is shown to remove these
extra states at the expense of losing manifest supersymmetry.
}

\np

\pagestyle{plain}
\pagenumbering{arabic}

\section{Introduction}{\label{S.1}}

Because of their conceptual simplicity, path-integral methods \ct{ref1,pol}
often provide convenient procedures to obtain insight in field theoretical
problems. In recent work by Strassler, McKeon, Schmidt, Schubert and others
\ct{strass}-\ct{vn1} world-line path integrals have been applied to a
reformulation of standard Feynman perturbation theory from which useful
information on the structure of perturbative Green's functions is extracted.
Some of these results were actually first derived in the particle-limit of
string theory \ct{bern}.

A basic question in this context is the representation of propagators in
quantum field theory by path integrals for relativistic particles of various
kind. In particular one would like to know the classical actions to be used
in these path-integrals, as well as the precise meaning of the functional
measure in some regularized form, e.g.\ by discretization. Answers to these
questions establish firm connections between the so-called first- and second
quantized formulations of relativistic quantum theory.

This paper addresses both of these problems. It pursues them from two
complementary points of view. First of all, there is the pragmatic question
of how to convert a given field-theory propagator to a path integral
expression. Such a procedure is discussed in this paper for both spin-0 and
spin-1/2 particles. Starting at the other end, one can also ask what kind of
quantum field theory is associated with a given classical action. The kind
of actions considered in this context are usually taken to possess some
desirable properties like reparametrization invariance and supersymmetry,
which pose however additional difficulties to quantization procedures since
gauge fixing then becomes necessary. Using previously established
BRST-procedures \ct{jw1,jw3} this is done in an extended state space with a
number of ghost- and auxiliary degrees of freedom. The propagators in these
extended state spaces are obtained, and it is then shown how to reduce them to
standard expressions in terms of physical variables only.

One result which deserves to be emphasized is that the irreducible
path-integral expression for the propagator of a massive Dirac fermion
is not based on a manifestly supersymmetric action, although
world-line supersymmetry is realized algebraically in the quantum theory
by the Dirac operator. Indeed, it is shown that the manifestly supersymmetric
version of the theory contains a `hidden' topological fermionic degree
of freedom which doubles the number of components of the physical states.
This theory therefore describes a degenerate doublet of fermions, rather
than a single massive Dirac particle. A path-integral expression for a
single massive Dirac fermion is also obtained (sect.\ref{S.7}). As in other
cases, the difference between the two models can be traced to the
representation of $\gam_5$.

This paper is organized as follows. In sects.(\ref{S.2})-(\ref{S.6}) the
free field theory of a scalar spin-0 particle is considered. A simple and
well-known path-integral expression is obtained, which is subsequently
rederived from the reparametrization-invariant classical model. These
calculations also help to explain the general procedures used.

In sect.(\ref{S.7}) a path-integral for a spin-1/2 particle is derived.
For massless particles it agrees with results in the literature \ct{cohen,hht}.
For massive fermions a new term is present in the action which has not
been considered before, and which is based on a bosonic representation of
$\gam_5$. The manifestly supersymmetric theory, which uses a fermionic
representation for $\gam_5$, is analyzed in sects.(\ref{S.8})-(\ref{S.10}),
and is shown to describe a doublet of spinors if the mass is non-zero.
Our conclusions are presented in sect.(\ref{S.11}).

\section{Free particle propagators}{\label{S.2}}

The Feynman propagator for a free  scalar particle of mass $m$ in
$D$-dimensional space-time is a specific solution of the inhomogeneous
Klein-Gordon equation

\be
\lh - \Box_x + m^2 \rh\, \Del_F (x - y)\, =\, \del^D(x - y),
\label{2.0}
\ee

\nit
such that positive frequencies propagate forward in time, and negative
frequencies backward. An explicit expression in terms of a Fourier
integral is

\be
\Del_{F}(x - y)\, =\, \int \frac{d^D p}{(2\pi)^D}\,
                     \frac{e^{i p \cdot (x-y)}}{p^2 + m^2 - i \ve} .
\label{2.1}
\ee

\nit
In the limit $\ve \rightarrow 0^+$ the simple real pole at positive $p^0 =
E(\vec{p}) = \sqrt{\vec{p}^{\,2} + m^2}$ gives the mass of the particle as
$m = E(0)$. The arbitrarily small imaginary part $i\ve$
implements the causality condition.

There is a straightforward connection between this propagator and the
classical mechanics of a relativistic point-particle through the
path-integral formalism \ct{ref1}. To establish this, let us first consider
the very general problem of finding the inverse of a non-singular hermitean
operator $\hH$. Following Schwinger \ct{ref2}, we construct the formal
solution

\be
\hH^{-1}\, =\, \lim_{\ve \rightarrow 0^+}\,
               i \int_0^{\infty} dT\, e^{-iT \lh \hH - i\ve \rh}.
\label{2.2}
\ee

\nit
Here the exponential operator

\be
\hK_{\ve}(T)\, =\, e^{-iT \lh \hH - i\ve \rh}
\label{2.3}
\ee

\nit
is the solution of the Schr\"{o}dinger equation

\be
i \dd{\hK_{\ve}(T)}{T}\, =\, \lh \hH - i\ve \rh\, \hK_{\ve}(T),
\label{2.4}
\ee

\nit
with the special properties\footnote{$\hI$ denotes the unit operator.}

\be
 \hK_{\ve}(0) = \hI, \hspace{3em}
\lim_{T \rightarrow \infty} \hK_{\ve}(T) = 0.
\label{2.5}
\ee

\nit
If the operator $\hH$ acts on the single-particle state-space with a complete
co-ordinate basis $\left\{ |x\rangle \right\}$, the matrix elements of the
operator in the co-ordinate basis are

\be
K_{\ve}(x - y|T)\, =\, \langle x| \hK_{\ve}(T) | y \rangle .
\label{2.6}
\ee

\nit
Completeness of the basis then implies Huygens' composition principle

\be
\int d^D x^{\prime}\, K_{\ve^{\prime}}(x - x^{\prime}|T^{\prime})\,
   K_{\ve^{\prime\prime}}(x^{\prime} - x^{\prime\prime}|T^{\prime\prime})\,
   =\, K_{\ve}(x - x^{\prime\prime}|T^{\prime} + T^{\prime\prime}).
\label{2.10}
\ee

\nit
Note that $\ve = (\ve^{\prime} T^{\prime} + \ve^{\prime\prime}
T^{\prime\prime})/(T^{\prime} + T^{\prime\prime})$ stays arbitrarily small if
$\ve^{\prime}$ and $\ve^{\prime\prime}$ are small enough.

Repeated use of eq.(\ref{2.10}) now allows one to write

\be
K_{\ve}(x-y|T)\, =\, \int \prod_{n=1}^N d^Dx_n \prod_{m=0}^N
    K_{\ve}\lh x_{m+1} - x_m | \Del T \rh,
\label{2.6.1}
\ee

\nit
with $\Del T = T/(N+1)$, and $x_0 = y$, $x_{N+1} = x$. Keeping $T$ fixed,
the limit $N \rightarrow \infty$ becomes an integral over continuous (but
generally non-differentiable) paths in co-ordinate space-time between points
$y$ and $x$. (Observe that $K_{\ve}(x-y|\Del T)$ depends only on the difference
$(x-y)$ and converges to $\del^D(x-y)$ for $\Del T \rightarrow 0$.)

If the operator $\hH$ is an ordered expression in terms of a canonical set of
operators $(\hx^{\mu},\hp_{\mu})$:

\be
\hH\, =\, \sum_{k,l}\, \hp_{\mu_1} ... \hp_{\mu_k} H^{\mu_1 ... \mu_k}_{\nu_1
          ... \nu_l} \hx^{\nu_1} ... \hx^{\nu_l},
\label{2.7}
\ee

\nit
then we can expand the co-ordinate path-integral expression (\ref{2.6.1})
further to a phase-space path-integral

\be
\ba{lll}
K_{\ve}(x - y|T) & = & \dsp{ \frac{1}{(2\pi)^D}\, \int d^D p_0
  \int \prod_{n=1}^{N} \frac{d^D x_n d^D p_n}{(2\pi)^D}\,
  e^{ i \sum_{k=0}^{N} \lh p_k \cdot (x_{k+1} - x_k)
  - \Del T H(p_k, x_k) \rh}  } \\
 & & \\
 & \rightarrow & \dsp{ \int_y^x \cD p(\tau) \cD x(\tau)\,
  e^{ i \int_0^T d\tau \lh p \cdot \dot{x} - H(p,x) \rh }. }
\ea
\label{2.8}
\ee

\nit
Here $H(p,x)$ is the c-number symbol of the ordered operator $\hH$, and
we have tacitly assumed that the ordered symbol of the exponential can be
replaced by the exponential of the ordered symbol. This is certainly correct
for the main applications we consider in this paper, as may be checked by
explicit calculations.

It is now clear, that one may interpret the symbol $H(p,x)$ as the hamiltonian
of some classical system, and the argument of the exponential as the
classical action. Integration over the momentum variables $p(\tau)$ then
in general leads to the lagrangian form of this action

\be
K_{\ve}(x - y|T)\, =\, \int_y^x \cD x(\tau)\,
  e^{i \int_0^T d\tau L(\dot{x}, x) },
\label{2.9}
\ee

\nit
where the precise meaning of the integration measure can be recovered
either from the phase-space expression (\ref{2.8}), or from requiring the
path-integral to satisfy Huygens' composition principle (\ref{2.10}).

Returning to eq.(\ref{2.0}), it states that $\Del_F(x - y)$ is the inverse
of the Klein-Gordon operator (in the space of square-integrable functions).
Rescaling it for later convenience by a factor $1/2m$, we consider the
evolution operator

\be
\hK_{\ve}(T)\, =\, \exp \left\{-\frac{iT}{2m} \lh -\Box + m^2 - i\ve \rh
     \right\}.
\label{2.10.1}
\ee

\nit
In the co-ordinate representation the explicit expression for the
matrix element of this operator is

\be
K_{\ve}(x-y|T)\, =\, -i \lh \frac{m}{2 \pi T}\rh^{D/2}\,
  e^{i \frac{m}{2T} (x-y)^2 - \frac{iT}{2} (m - i \ve)}.
\label{2.10.2}
\ee

\nit
The Feynman propagator can then be written as

\be
\Del_F(x-y)\, =\, \frac{i}{2m}\, \int_0^{\infty} dT K_{\ve}(x-y|T),
\label{2.10.3}
\ee

\nit
and using the previous results it can be cast in the form of a path integral
\ct{cas}

\be
\Del_F(x - y)\, =\,  \frac{i}{2m}\, \int_0^{\infty} dT\, \int_y^x \cD
      x(\tau)\, \exp \left\{\frac{im}{2}\,
      \int_0^T d\tau \lh \dot{x}_{\mu}^2 - 1 \rh \right\}.
\label{2.11}
\ee

\nit
The same expression is also obtained directly by iteration of (\ref{2.10.2})
using the product formula (\ref{2.6.1}). For massless particles yet another
derivation, based on the concepts of moduli space, has been discussed in
\ct{cohen,hht}. The result shows, that the scalar propagator is connected
to a classical particle model with lagrangian

\be
L = \frac{m}{2}\, \lh \dot{x}_{\mu}^2 - 1 \rh ,
\label{2.11.1}
\ee

\nit
related by Legendre transform to the simple hamiltonian

\be
H(p,x)\, =\, \frac{p_{\mu}^2 + m^2}{2m}.
\label{2.12}
\ee

\nit
Because $H(p,x)$ is quadratic in $p_{\mu}$ and independend of $x^{\mu}$,
the integration measure in the path integral (\ref{2.11}) is just the free
gaussian measure of ref.\ct{ref1}.

The path-integral representation of the propagator establishes a simple
connection between scalar quantum field theory and the classical point
particle model (\ref{2.11.1}). As is well-known, this connection does not
only hold at the level of the action or hamiltonian, it also extends to the
dynamics, in the sense that paths in the neighborhood of the solutions of the
classical equations of motion give the dominant contribution to the path
integral, certainly for the simple model discussed where the path integral is
a pure Gaussian.

\section{Reparametrization invariance}{\label{S.3}}

The classical equations of motion which follow from the lagrangian
(\ref{2.11.1}) state that the momentum is constant along the
particle worldline:

\be
\dot{p}^{\mu} = m \ddot{x} = 0,
\label{3.1}
\ee

\nit
Therefore this classical theory seems to contain less information
than the quantum theory from which we started: we have to recover
the condition that the momentum should lie on the mass shell:

\be
p^2_{\mu} + m^2 = 0.
\label{3.2}
\ee

\nit
This condition is equivalent to the vanishing of the hamiltonian
(\ref{2.12}). Since the hamiltonian is the generator of translations
in the worldline parameter $\tau$, the mass-shell condition is
recovered by requiring the dynamics of the particle to be independent
of the world-line parametrization.

As is well-known \ct{brink1,brink2}, this can be achieved by introducing a
gauge variable $e(\tau)$ for re\-par\-am\-etri\-za\-tions of the world line
(the einbein). Under local reparametrizations $\tau \rightarrow \tau^{\prime}
= f(\tau)$, let the co-ordinates and einbein transform as

\be
x^{\mu}(\tau) \rightarrow x^{\prime \mu} (\tau^{\prime}) = x^{\mu}(\tau),
\hspace{3em}
e(\tau) \rightarrow e^{\prime}(\tau^{\prime})
        = e(\tau) \frac{d\tau}{d\tau^{\prime}}.
\label{3.3}
\ee

\nit
Then a reparametrization-invariant action can be constructed of the form

\be
S_{cl}[x^{\mu}(\tau),e(\tau)]\, =\, \frac{m}{2}\, \int_0^T d\tau\,
 \lh \frac{1}{e}\, \dot{x}_{\mu}^2 - e \rh.
\label{3.4}
\ee

\nit
The equations of motion one derives from this action are

\be
\frac{1}{e} \frac{d}{d\tau}\, \frac{1}{e} \frac{dx^{\mu}}{d\tau}\, =\, 0,
\hspace{3em} \lh \frac{1}{e} \frac{dx^{\mu}}{d\tau} \rh^2\, +\, 1\, =\, 0.
\label{3.5}
\ee

\nit
Identifying $d\tilde{\tau} = e d\tau$ with the proper time, and as a
consequence $p^{\mu} = m dx^{\mu}/d\tilde{\tau}$ with the proper momentum,
this reproduces both the world-line equation of motion and the mass-shell
condition.

For the new hamiltonian we can take

\be
H = \frac{e}{2m}\, \lh p_{\mu}^2 + m^2 \rh .
\label{3.6}
\ee

\nit
This generates proper-time translations on special phase-space functions
$F(x(\tau),p(\tau))$ by the Poisson-brackets:

\be
\frac{dF}{d\tilde{\tau}}\, =\, \frac{1}{e} \frac{dF}{d\tau}\,
                           =\, \frac{1}{e}\, \left\{ F, H \right\} .
\label{3.7}
\ee

\nit
Note, that one cannot impose the constraint that $H$ vanishes before
we compute the brackets.

The difficulty with this formulation is evidently the additional dynamical
variable $e(\tau)$, for which the evolution is not fixed by the Euler-Lagrange
equations derived from $S_{cl}[x^{\mu},e]$, and which has no conjugate
momentum. The origin of these difficulties is precisely the reparametrization
invariance (\ref{3.3}). As a result, the hamiltonian evolution equation
(\ref{3.7}) does not hold for arbitrary functions on the {\em complete} phase
space, $F(x^{\mu},p^{\mu},e)$. Clearly, to recover the results of
sect.\ref{S.2} it is necessary to fix $e(\tau)$ to a constant value and change
$\tau \rightarrow \tilde{\tau}$, which amounts to a rescaling of the unit of
time on the worldline such that the particle's internal clock and the
laboratory clock tick at the equal rates when the particle is at rest.

Therefore one chooses a gauge

\be
\dot{e} = 0,
\label{3.8}
\ee

\nit
implying that $e$ is constant, and adds this condition to the equations of
motion. But this can only be done after the variation of $e$ in the
action has produced the mass-shell constraint (\ref{3.2}). It is
therefore of interest to have a formalism in which one can impose the
gauge condition (\ref{3.8}) from the start, and still keep track of all the
constraints imposed by the reparametrization invariance (\ref{3.3}). An
appropriate formalism to solve this problem is the BRST procedure (for an
introduction, see for example refs.\ct{KO,henn,jw2}), which we describe here
for the case at hand \ct{jw1}.

\section{BRST formulation}{\label{S.4}}

We impose the gauge condition (\ref{3.8}) by adding it to the action with
a Lagrange multiplier $\lb$; at the same time we introduce a corresponding
Faddeev-Popov ghost action, using (real) anti-commuting ghost variables
$(b,c)$, in such a way that the complete action is invariant under the
Grassmann-odd, nilpotent BRST transformations

\be
\ba{llllll}
\dL x^{\mu} & = c \dot{x}^{\mu},       \hspace{2em} &
      \dL e & = c \dot{e} + \dot{c} e, \hspace{2em} & \dL c & = c \dot{c}, \\
 & & & & & \\
\dL b & = - i \lb, & \dL \lb & = 0.
\ea
\label{3.9}
\ee

\nit
After a convenient rescaling $ec \rightarrow c$, turning the ghost $c$ into a
world-line scalar density such that $\dL e = \dot{c}$ and $\dL c = 0$, the
BRST-invariant gauge fixed action reads

\be
S_{gf}\, =\, S_{cl}[x(\tau), e(\tau)]\, +\, \int_0^T d\tau\,
             \lh \lb \dot{e} + i \dot{b} \dot{c} \rh.
\label{3.10}
\ee

\nit
Actually, because a partial integration has been performed in the ghost term,
the action is invariant only modulo a total time derivative. This is
sufficient. Note, that the Lagrange multiplier $\lb$ now plays the role of
momentum $p_e$ conjugate to the einbein $e$.

The action $S_{gf}$ has no local invariances left and can be treated by the
usual procedures of canonical hamiltonian analysis. The canonical momenta
are

\be
\ba{llll}
p^{\mu} & \dsp{ = \frac{m}{e}\, \dot{x}^{\mu}, } \hspace{2em}
        & p_e & = \lb, \\
 & & & \\
\pi_c & = - i \dot{b}, & \pi_b & = i \dot{c}.
\ea
\label{3.11}
\ee

\nit
Note that the ghost-momenta $(\pi_c, \pi_b)$ are imaginary.
The gauge-fixed hamiltonian is

\be
H_{gf}\, =\, \frac{e}{2m}\, \lh p_{\mu}^2 + m^2 \rh - i \pi_b \pi_c.
\label{3.12}
\ee

\nit
The equations of motion are given by the Poisson brackets

\be
\frac{dF}{d\tau}\, =\, \left\{F, H_{gf} \right\},
\label{3.13}
\ee

\nit
where the brackets are defined by

\be
\ba{lll}
\left\{F, G \right\} & = & \dsp{ \dd{F}{x^{\mu}} \dd{G}{p_{\mu}}\, -\,
     \dd{F}{p_{\mu}} \dd{G}{x^{\mu}}\, +\, \dd{F}{e} \dd{G}{p_e}\, -\,
     \dd{F}{p_e} \dd{G}{e} } \\
 & & \\
 & & \dsp{ +\, (-1)^{a_F} \lh \dd{F}{c} \dd{G}{\pi_c} + \dd{F}{\pi_c} \dd{G}{c}
        \, +\, \dd{F}{b} \dd{G}{\pi_b} + \dd{F}{\pi_b} \dd{G}{b} \rh. }
\ea
\label{3.14}
\ee

\nit
Here all derivatives are taken from the left, and $a_F$ is the Grassmann parity
of $F$.

To recover the constraints imposed by local reparametrization invariance,
one constructs the (Grassmann-odd) conserved BRST charge

\be
\Og\, =\, \frac{c}{2m}\, \lh p_{\mu}^2 + m^2 \rh\, -\, i \pi_b p_e,
\hspace{3em} \left\{ \Og, H_{gf} \right\}\, =\, 0.
\label{3.15}
\ee

\nit
This BRST charge is nilpotent in the sense that

\be
\left\{ \Og, \Og \right\}\, =\, 0.
\label{3.15.1}
\ee

\nit
The BRST principle makes use of this property, and states that the physical
observables of the theory are the cohomology classes of the BRST operator
on the phase-space functions $F(x,p;e,p_e;c,\pi_c;b,\pi_b)$ with ghost
number\footnote{We assign ghost number +1 to $(c,\pi_b)$, and -1 to
$(b,\pi_c)$.} $N_{gh} = 0$ \ct{ff,henn2}. More precisely, physical
quantities are BRST invariant:

\be
\left\{ F, \Og \right\}\, =\, 0,
\label{3.16}
\ee

\nit
but this allows an ambiguity in $F$ as a result of (\ref{3.15.1});
this ambiguity is resolved by associating obervables $F$ with equivalence
classes of functions differing only by a BRST transformation:

\be
F^{\prime}\, \sim\, F \hspace{2em} \Leftrightarrow \hspace{2em}
F^{\prime}\, =\, F\, +\, \left\{ \Lb, \Og \right\},
\label{3.17}
\ee

\nit
with $\Lb$ a phase-space function of Grassmann parity opposite to $F$,
and one unit in ghost-number lower. In particular, any function which can be
written as a pure BRST transform

\be
G\, =\, \left\{ \Lb, \Og \right\},
\label{3.18}
\ee

\nit
is in the equivalence class of zero and may be taken to vanish. In the case
of the relativistic particle, this happens to be true for the hamiltonian
itself:

\be
H_{gf}\, =\, - \left\{ e\pi_c, \Og \right\},
\label{3.19}
\ee

\nit
and for the classical hamiltonian as well:

\be
p_{\mu}^2\, +\, m^2\, =\, -2m \left\{ \pi_c, \Og \right\}.
\label{3.20}
\ee

\nit
Hence the classical and ghost terms in the hamiltonian $H_{gf}$ vanish
separately and the mass-shell condition follows. With

\be
p_e\, =\, -i\, \left\{ b, \Og \right\},
\label{3.21}
\ee

\nit
we also recover the vanishing of the momentum conjugate to the einbein $e$.

Before turning to the quantum theory, we draw attention to a peculiarity of
the action (\ref{3.10}): it is invariant under SO(2) rotations in the ghost
variables $(b,c)$. Therefore this theory possesses a second BRST invariance
with conserved nilpotent charge

\be
\tilde{\Og}\, =\,\frac{b}{2m}\, \lh p_{\mu}^2 + m^2 \rh\, -\, i \pi_c p_e,
\hspace{3em} \left\{ \tilde{\Og}, H_{gf} \right\}\, =\, 0.
\label{3.22}
\ee

\nit
The algebra of these two BRST charges has the property that

\be
\left\{ \Og, \tilde{\Og} \right\}\, =\,
        \frac{i}{m}\, \lh p_{\mu}^2 + m^2 \rh\, p_e.
\label{3.23}
\ee

\nit
Note that both factors on the right-hand side are separately conserved,
and vanish on the physical hyperplane in phase space.

\section{BRST quantization}{\label{S.5}}

Having formulated a complete, BRST-invariant (pseudo-)classical mechanics
for the relativistic point particle, we now return to the quantum theory and
study the relation between this model and quantum field theory. Various
aspects of this problem have been discussed in \ct{henn3,mon,jw1}. To obtain
additional insight, we construct a quantum theory corresponding to the
classical hamiltonian (\ref{3.12}) in an extended state-space, compute the
propagator and establish the relation with the usual Feynman propagator
(\ref{2.1}), thereby showing the physical equivalence of these different
formulations of scalar field theory.

The quantum theory of interest is obtained by replacing the
phase-space variables $(x^{\mu}, p_{\mu}; e, p_e; c, \pi_c; b, \pi_b)$ by
operators, and postulating (anti-)\-commutation relations between them in
direct correspondence with the Poisson-brackets (\ref{3.14}):

\be
\ba{llll}
\left[ \hx^{\mu}, \hp_{\nu} \right] & =\; i \del^{\mu}_{\nu}, &
\left[ \he, \hp_e \right] & =\; i, \\
 & & & \\
\left\{ \hc, \hat{\pi}_c \right\} & =\; - i, &
\left\{ \hb, \hat{\pi}_b \right\} & =\; - i.
\ea
\label{5.1}
\ee

\nit
In the co-ordinate representation these algebraic relations hold on
making the identification

\be
\ba{llll}
\hp_{\mu} & \dsp{ =\, - i\dd{}{x^{\mu}},} & \hp_e & \dsp{ =\, - i\dd{}{e}, }\\
 & & & \\
\hat{\pi}_c & \dsp{ =\, - i \dd{}{c},} & \hat{\pi}_b & \dsp{ =\, -i\dd{}{b}.}
\ea
\label{5.2}
\ee

\nit
The bosonic momenta are self adjoint, and the fermionic ones
anti self-adjoint with respect to the inner product

\be
\lh \Phi, \Psi \rh\, =\, i \int d^Dx^{\mu} de \int db dc\, \Phi^* \Psi,
\label{5.2.1}
\ee

\nit
where wave functions like $\Psi$ are polynomials in the ghost variables,
with co-efficients depending on the remaining co-ordinates $(x^{\mu},e)$:

\be
\Psi(b,c)\, =\, \psi - ib \psi_b + c \psi_c - icb \psi_{cb}.
\label{5.2.2}
\ee

\nit
The inner product (\ref{5.2.1}) then reads in components

\be
\lh \Phi, \Psi \rh\, =\, \phi^* \psi_{cb} + \phi^*_b \psi_c + \phi^*_c \psi_b
                         + \phi^*_{cb} \psi.
\label{5.2.3}
\ee

\nit
Clearly this form is not positive definite. However, with respect to
this inner product the hamiltonian operator,

\be
\hH_{gf}\, =\, \frac{e}{2m}\, \lh - \Box + m^2 \rh\, +\,
               i \frac{\pl^2}{\pl b\pl c},
\label{5.3}
\ee

\nit
and the nilpotent BRST operator,

\be
\hOg\, =\, \frac{c}{2m}\, \lh - \Box + m^2 \rh\, +\,
           i \frac{\pl^2}{\pl e \pl b}, \hspace{3em}
\hOg^2\, =\, 0,
\label{5.4}
\ee

\nit
are self adjoint. The indefinite metric implicit in this inner product
is a general and necessary feature of a model with a nilpotent self-adjoint
operator like $\hOg$ \ct{jw3}.

It is also possible to define a positive-definite inner product on the
state space: introduce a duality operation \ct{jw6}

\be
\Psi\, \rightarrow\, \tilde{\Psi}\, =\, \psi_{cb} - ib \psi_c + c \psi_b
       - icb \psi,
\label{5.5}
\ee

\nit
and define

\be
\ba{lll}
\langle \Phi, \Psi \rangle & = & \lh \Phi, \tilde{\Psi} \rh \\
 & & \\
 & = & \phi^* \psi + \phi^*_b \psi_b + \phi^*_c \psi_c + \phi^*_{cb} \psi_{cb}.
\ea
\label{5.6}
\ee

\nit
With respect to this positive definite inner product the fermionic momenta
and the BRST charge are no longer (anti-)self adjoint. In particular $\hc^*
= i \hat{\pi}_c$, $\hb^* = i \hat{\pi}_b$, and

\be
\langle \hOg \Phi, \Psi \rangle\, =\, \langle \Phi, \hOg^* \Psi \rangle,
\label{5.7}
\ee

\nit
where the adjoint BRST charge, also known as the co-BRST operator, is

\be
\hOg^*\, =\, \frac{1}{2m}\, \lh - \Box + m^2 \rh\, \dd{}{c}\,
             +\, ib \dd{}{e}.
\label{5.8}
\ee

\nit
Like the BRST charge itself, $\hOg^*$ is nilpotent; however, it is not
conserved:

\be
\left[ \hOg^*, \hH_{gf} \right]\, =\, \tilde{\Og},
\label{5.9}
\ee

\nit
where $\tilde{\Og}$ is the operator corresponding to the dual BRST charge
we have encountered before in (\ref{3.22}):

\be
\tilde{\Og}\, =\,
       \frac{ib}{2m} \lh -\Box + m^2 \rh\, +\, \frac{\pl^2}{\pl e \pl c}.
\label{5.10}
\ee

\nit
We also note in passing, that the BRST operator $\hOg$ itself is obtained
from the commutator of the hamiltonian with the adjoint of the
dual BRST charge:

\be
\left[ \tilde{\Og}^*, \hH_{gf} \right]\, =\, - \hOg,
\label{5.11}
\ee

\nit
with

\be
\tilde{\Og}^*\, =\, - \frac{i}{2m}\, \lh -\Box + m^2 \rh\, \dd{}{b}\,
                    -\, c \dd{}{e}.
\label{5.12}
\ee

\nit
We can now show how the physical states of the scalar particle are reobtained
in the BRST formalism through the cohomology of $\hOg$. One identifies the
physical states with equivalence classes of states which are BRST invariant
and differ only by a BRST-exact term:

\be
\hOg\, \Psi_{phys}\, =\, 0, \hspace{3em}
\Psi_{phys}\, \sim\, \Psi^{\prime}_{phys}\, =\,
                     \Psi_{phys}\, +\, \hOg\, \Lambda.
\label{5.13}
\ee

\nit
To obtain exactly one representative of each BRST invariance class,
consider the zero-modes of the BRST laplacian \ct{jw3,jw1}

\be
\ba{lll}
\Del_{BRST} & = & \hOg \hOg^*\, +\, \hOg^* \hOg \\
 & & \\
 & = & \dsp{
       \frac{1}{4m^2}\, \lh -\Box + m^2 \rh^2\, -\, \frac{\pl^2}{\pl e^2}. }
\ea
\label{5.14}
\ee

\nit
This operator is positive definite, and its zero-modes are zero-modes
of the two terms on the right-hand side separately. Therefore

\be
\Del_{BRST}\, \Psi_{phys}\, =\, 0 \hspace{1em} \Leftrightarrow \hspace{1em}
\left\{ \lh -\Box + m^2 \rh\, \Psi_{phys}\, =\, 0\, \wedge\,
        \dd{}{e} \Psi_{phys}\, =\, 0 \right\}.
\label{5.15}
\ee

\nit
Hence the wave functions $\Psi_{phys}$ indeed satisfy the Klein-Gordon
equation and are independend of the gauge variable $e$.

\section{Gauge-fixed propagator and path integral}{\label{S.6}}

The next step is to construct the propagator for the scalar particle
as described by the wave functions (\ref{5.2.2}) in the BRST-extended
state space, using the formalism of sect.(\ref{S.2}). Let us label the
co-ordinates $(x^{\mu},e,b,c)$ collectively by $Z$. Then the
Schr\"{o}dinger equation for the kernel $K_{gf}(Z;Z^{\prime}|T)$ in
the co-ordinate picture becomes:

\be
i \dd{}{T}K_{gf}(Z;Z^{\prime}|T)\, =\, \lh \frac{e}{2m} \lh -\Box +
  m^2 - i\ve \rh + i \frac{\pl^2}{\pl b\pl c}  \rh\, K_{gf}(Z;Z^{\prime}|T).
\label{6.1}
\ee

\nit
This kernel is required to satisfy the initial condition

\be
\ba{lll}
K_{gf}(Z;Z^{\prime}|0) & = & \del(Z - Z^{\prime}) \\
 & & \\
 & = &  \del^D(x-x^{\prime}) \del(e-e^{\prime}) \del(c-c^{\prime})
        \del(b-b^{\prime}).
\ea
\label{6.2}
\ee

\nit
The solution to this equation is

\be
K_{gf}(Z;Z^{\prime}|T)\, =\,
 -iT \del(e-e^{\prime})\, \lh \frac{m}{2\pi eT} \rh^{D/2}\,
 e^{i \left[ \frac{m}{2eT} \lh x-x^{\prime} \rh^2 - \frac{eT}{2}\lh m - i\ve
\rh
     + \frac{i}{T} \lh b-b^{\prime} \rh \lh c-c^{\prime} \rh \right]}.
\label{6.3}
\ee

\nit
This is the direct counterpart of eq.(\ref{2.10.2}). It is straightforward
to verify that the initial condition (\ref{6.2}) is satisfied, as is the
composition principle:

\be
\int dZ^{\prime}\, K_{gf}(Z;Z^{\prime}|T^{\prime})\,
   K_{gf}(Z^{\prime};Z^{\prime\prime}|T^{\prime\prime})\, =\,
   K_{gf}(Z;Z^{\prime\prime}|T^{\prime} + T^{\prime\prime}),
\label{6.4}
\ee

\nit
where the integration measure reads

\be
\int dZ\, =\, i \int d^D x de \int db dc.
\label{6.4.1}
\ee

\nit
The propagator in the extended state space then becomes

\be
\Del_{gf}(Z;Z^{\prime})\, =\, \frac{ie}{2m}\, \int_{0}^{\infty} dT\,
                              K_{gf}(Z;Z^{\prime}|T).
\label{6.5}
\ee

\nit
It is the solution of the inhomogeneous extended Klein-Gordon equation

\be
\lh - \Box + m^2 - i\ve + \frac{2im}{e} \frac{\pl^2}{\pl b\pl c} \rh\,
     \Del_{gf}(Z;Z^{\prime})\, =\, \del(Z - Z^{\prime}).
\label{6.6}
\ee

\nit
Substitution of the expression (\ref{6.3}) for the kernel
$K_{gf}(Z,Z^{\prime}|T)$ in eq.(\ref{6.5}) for the generalized propagator
gives a closed expression for the propagator in co-ordinate space which is
conveniently represented in terms of a momentum integral.
Switching to the Fourier representation of the co-ordinate part and
performing the integral over $T$ yields

\be
\Del_{gf}(x,e,b,c;x^{\prime},e^{\prime},b^{\prime},c^{\prime})\, =\,
  \frac{2m}{ie}\, \del \lh e - e^{\prime} \rh\, \int \frac{d^D p}{(2\pi)^D}\,
  e^{i p \cdot (x - x^{\prime})}\, \lh \frac{ \dsp{
  e^{- \frac{ie}{2m} \lh p^2 + m^2 - i \ve \rh \lh b - b^{\prime} \rh
     \lh c - c^{\prime} \rh } }}{ \left[ p^2 + m^2 - i \ve \right]^2 } \rh .
\label{6.9}
\ee

\nit
{}From this expression it is straightforward to obtain the Feynman propagator
by integration over the unphysical degrees of freedom:

\be
\ba{lll}
\Del_F(x-x^{\prime}) & = & \dsp{ \int_{\infty}^{\infty} de \int db \int dc\,
   \Del_{gf}(x,e,b,c;x^{\prime},e^{\prime},b^{\prime},c^{\prime}) } \\
 & & \\
 & = & \dsp{ \int \frac{d^Dp}{(2\pi)^D}\,
       \frac{ e^{i p \cdot (x - x^{\prime})} }{p^2 + m^2 - i \ve}.}
\ea
\label{6.10}
\ee

\nit
Alternatively, repeated use of the composition principle can be used to
construct a con\-fig\-ura\-tion space path integral representation of the
propagator. Introducing the Fourier representation for the delta function
$\del(e-e^{\prime})$ repeated use of eq.(\ref{6.4}) gives in explicit
notation:

\be
\Del_{gf}(x,e,b,c;x^{\prime},e^{\prime},b^{\prime},c^{\prime})\, =\,
  \frac{ie}{2m}\, \int_0^{\infty} dT \int_{\Gam} \left[ \cD x(\tau)
  \cD \lb(\tau) \cD e(\tau) \cD c(\tau) \cD b(\tau) \right]\,
  e^{ i S_{gf}[x,\lb,e,b,c] },
\label{6.7}
\ee

\nit
where $S_{gf}$ is the classical gauge-fixed action (\ref{3.10}) and the
functional integral is over all paths $\Gam$ in the configuration space
between $(x^{\prime},e^{\prime},b^{\prime},c^{\prime})$ and $(x,e,b,c)$.
{}From the construction a consistent discrete regularization,
specifying the measure to be used, is obtained:

\be
\ba{l}
\dsp{ \int_{\Gam} \left[ \cD x(\tau) \cD \lb(\tau) \cD e(\tau) \cD c(\tau)
      \cD b(\tau) \right]\, e^{ i S_{gf}[x,\lb,e,b,c] }\, = }\\
  \\
  \dsp{ \hspace{2em}  =\, \lim_{N \rightarrow \infty}\,
      \int \prod_{n=1}^{N} \left[ d^D x_n d \lb_n de_n d c_n d b_n\,
       \frac{i\Del T}{2\pi}\, \lh \frac{m}{2\pi i e_n \Del T} \rh^{D/2} \right]
       \times } \\
 \\
\dsp{ \hspace{2em} e^{ i \sum_{k=0}^N \Del T \left[ \frac{m}{2e_k}
      \lh \frac{x^{\mu}_{k+1} - x^{\mu}_k}{\Del T} \rh^2 - \frac{m}{2} e_k
      + \lb_{k+1} \frac{\lh e_{k+1} - e_k \rh}{\Del T} + i \lh \frac{b_{k+1}
      - b_k}{\Del T} \rh  \lh \frac{c_{k+1} - c_k}{\Del T} \rh \right] }, }
\ea
\label{6.8}
\ee

\nit
where again $\Del T = T/(N+1)$, with $T$ fixed. Note that the measure
contains a factor $\sqrt{m/2\pi i e \Del T}$ for each integral over a
co-ordinate $x^{\mu}$, and a factor $\sqrt{i\Del T}$ for integration over a
ghost $b$ or $c$. Eqs.(\ref{6.7}) and (\ref{6.8}) give a precise meaning to
relation between the path-integral representation of the propagator in the
extended state space and the BRST-invariant gauge-fixed classical action
(\ref{3.10}).

\section{Fermions}{\label{S.7}}

The Dirac-Feynman propagator for a free fermion is the solution of the
inhomogeneous Dirac equation

\be
\lh \gam \cdot \pl_x + m \rh \, S_F(x - y)\, =\, \del^D(x-y),
\label{7.1}
\ee

\nit
with the same causal boundary conditions as for the scalar particle. The
Dirac matrices $\gam^{\mu}$ form a $D$-dimensional Clifford algebra and are
normalized to satisfy the anti-commutation relations

\be
\left\{ \gam^{\mu}, \gam^{\nu} \right\}\, =\, 2\, \eta^{\mu\nu}.
\label{7.2}
\ee

\nit
The standard Fourier integral representation of the Dirac-Feynman propagator is

\be
S_F(x-y)\, =\, \int \frac{d^D p}{(2\pi)^D}\, \frac{ \lh - i \gam \cdot p
               + m \rh } {p^2 + m^2 - i \ve}\, e^{ i p \cdot (x - y)}.
\label{7.3}
\ee

\nit
It is possible to use the Schwinger procedure described in
sect.\ref{S.2} to construct a path-integral representation of this
propagator as a Clifford-algebra valued object. There is however
another method that is often preferred in applications, especially
when interactions are introduced into the model. This method uses
anti-commuting variables to represent the Clifford algebra \ct{Ber,Fad,BDW};
the application to spinning particles has been studied for example in
refs.\ \ct{hht,brink1,brink2}, \ct{BerMar}-\ct{jw4}. The use of this
method here allows a straightforward path-integral representation
of the Feynman propagator for fermions in terms of bosonic co-ordinates
$x^{\mu}$ and a matching set of fermionic (Grassmann-odd) co-ordinates
$\psi^{\mu}$. However, as the details of the procedure depend on the
number of dimensions, we will from now on choose $D = 4$, and limit
ourselves to that physically relevant case. Modifications to treat
the same problem in another number of dimensions are straightforward
to make.

In order to achieve the transition to anti-commuting variables, we define
a representation of the Clifford algebra in terms of two anti-commuting
variables $(\xi^1, \xi^2)$ by

\be
\ba{ll}
\dsp{ \hgm^1 \,=\, \xi^1 + \dd{}{\xi^1} , } &
\dsp{ \hgm^2 \,=\, -i \lh \xi^1 - \dd{}{\xi^1} \rh, } \\
 & \\
\dsp{ \hgm^3 \,=\, \xi^2 + \dd{}{\xi^2} , } &
\dsp{ \hgm^0 \,=\, -i \hgm^4\, =\, \xi^2 - \dd{}{\xi^2} . }
\ea
\label{7.3.1}
\ee

\nit
In addition to the $\hgm^{\mu}$ we also need the usual pseudo-scalar element
of the Clifford algebra:

\be
\ba{lll}
\hgm_5 & = & \dsp{ - \frac{i}{4!}\, \ve_{\mu\nu\kg\lb}
             \hgm^{\mu} \hgm^{\nu} \hgm^{\kg} \hgm^{\lb} } \\
 & & \\
 & = & \dsp{ \lh 2 \xi^1 \dd{}{\xi^1} - 1 \rh
             \lh 2 \xi^2 \dd{}{\xi^2} - 1 \rh . }
\ea
\label{7.3.3}
\ee

\nit
Note that in contrast to the other $\hgm^{\mu}$, in this realization
$\hgm_5$ is represented by a Grassmann-even (bosonic) operator. Therefore
we refer to this representation as the {\em bosonic} form of the $\hgm_5$.
In later sections we will also encounter a representation of $\hgm_5$
in terms of a Grassmann-odd operator, which is appropriately refered to
as the {\em fermionic} representation.
Now the following algebraic relations are satisfied by these operators:

\be
\left\{ \hgm^{\mu}, \hgm^{\nu} \right\}\, =\, 2 \eta^{\mu\nu} ,
\label{7.3.2}
\ee

\nit
and

\be
\hgm_5^2\, =\, 1 , \hspace{3em}
\left\{ \hgm_5, \hgm^{\mu} \right\}\, =\, 0.
\label{7.3.4}
\ee

\nit
These operators therefore realize the Clifford algebra of the Dirac matrices
including $\gam_5$.

The operators $(\hgm ^{\mu},\hgm _5)$ act on spinors $\Phi(\xi^1,\xi^2)$,
here defined as functions with a Grassmann polynomial structure

\be
\Fg (\xi^1,\xi^2)\, =\, \fg_2\, +\, \xi^1 \fg_3\, -\, \xi^2 \fg_4\, -\,
     \xi^1 \xi^2 \fg_1,
\label{7.4}
\ee

\nit
where all co-efficients are functions either of co-ordinates or momentum.
One can define the usual positive-definite inner product for spinors:

\be
\langle \Phi, \Psi \rangle\, =\, \sum_{\ag}\, \fg^*_{\ag} \psi_{\ag}.
\label{7.3.5}
\ee

\nit
This inner product can be written in the Grassmann representation as

\be
\langle \Phi, \Psi \rangle\, =\, \int \prod_{k} \lh d\xi^k d\bar{\xi}^k \rh\,
   e^{\bar{\xi} \cdot \xi}\, \Phi^*(\bar{\xi}) \Psi (\xi),
\label{7.3.5.1}
\ee

\nit
where a second set of Grassmann variables $\bar{\xi}^{1,2}$ has been
introduced as argument of the conjugate wave function, and the star
denotes complex conjugation of the c-number co-efficients of $\Phi (\xi)$
plus reversal of the order of the $\xi^k$. It is straightforward to
check that w.r.t.\ this inner product

\be
\langle \Phi, \dd{}{\xi^k} \Psi \rangle\, =\,
   \langle \xi^k \Phi, \Psi \rangle.
\label{7.3.5.2}
\ee

\nit
This result implies that in contrast to the other $\hgm^i$ $(i = 1,2,3)$ and
$\hgm^4$, the time-like operator $\hgm^0$ is not real w.r.t.\ $\langle \Phi,
\Psi \rangle $. This is of course to be expected from the lorentzian signature
of space-time. On the other hand, hermiticity can be restored by defining a
(lorentz invariant) indefinite-metric scalar product $\langle \overline{\Phi},
\Psi \rangle $ using the Pauli-conjugate spinorial wave function

\be
\overline{\Fg}(\xi^1,\xi^2)\, =\, \fg_4^*\, +\, \xi^1 \fg_1^*\, -\,
                             \xi^2 \fg_2^*\, -\, \xi^1 \xi^2 \fg_3^*.
\label{7.4.2}
\ee

\nit
In general the co-efficients of $\Fg$ are independent complex numbers, in
which case they represent the components of a Dirac spinor $\fg_{\ag}$.
Irreducible Weyl spinors can be obtained as eigenfunctions of $\hgm_5$.
As

\be
\hgm_5\, \Fg(\xi^1,\xi^2)\, =\, \Fg(-\xi^1,-\xi^2) ,
\label{7.4.1}
\ee

\nit
it follows that the Grassmann-even components define a Weyl spinor
of positive chirality, whilst the Grassmann-odd components define
a Weyl spinor of negative chirality. In the representation chosen
here the chirality can therefore be identified with the Grassmann
parity of the spinor $\Fg$.

It is also possible to represent Majorana spinors by requiring $C \Fg =
\bar{\Fg}$, where $\bar{\Fg}$ is the Pauli-conjugate spinor
and $C$ is the (anti-hermitean) charge-conjugation operator

\be
C\, =\, \lh \xi^1 - \dd{}{\xi^1} \rh\, \lh \xi^2 - \dd{}{\xi^2} \rh.
\label{7.4.3}
\ee

\nit
The Majorana constraint results in the component relations $\fg_4 =
\fg_1^*$ and $\fg_3 = - \fg_2^*$. In the following we consider Dirac
spinors unless explicitly stated otherwise.

With the above definitions, the Dirac equation can now be transcribed
as follows: let

\be
\Fg^{\prime}(\xi^1,\xi^2)\, =\,
     \lh \hgm \cdot p + m \hgm_5 \rh\, \Fg(\xi^1,\xi^2);
\label{7.5}
\ee

\nit
then the components of $\Fg^{\prime}$ are then related to those of $\Fg$ by

\be
\fg^{\prime}_{\ag}\, =\, \left[ \lh - i \gam \cdot p + m \rh \, \gam_5\,
   \right]_{\ag\bg}\, \fg_{\bg},
\label{7.6}
\ee

\nit
in the representation of the Dirac matrices defined by

\be
\ba{ll}
\dsp{ \gam_i\, =\, \left( \ba{cc}
                          0 & -i \sg_i \\
                          i \sg_i & 0
                          \ea \right),    } & \dsp{
                                                \gam^4\, =\, i \gam^0\, =\,
                                                \left( \ba{cc}
                                                   0 & 1 \\
                                                   1 & 0
                                                   \ea \right), } \\
 & \\
\dsp{ \gam_5\, =\, \left( \ba{cc}
                      1 & 0 \\
                      0 & -1
                      \ea \right). }
\ea
\label{7.7}
\ee

\nit
Since the matrix $\gam_5$ is unitary, the equation $\Fg^{\prime} = 0$
is completely equivalent with the free Dirac equation in momentum space,
and we infer that $S_F(x-y)\gam_5$ can be identified with the inverse of
the Dirac operator $(-i \hgm \cdot \pl + m \hgm_5)$ in co-ordinate space.
In the process of translation we have obtained the following correspondence

\be
\hgm^{\mu}\, \mapsto\, -i \gam^{\mu} \gam_5, \hspace{3em}
\hgm_5\, \mapsto \gam_5.
\label{7.8}
\ee

\nit
The next step is to write the inverse of the Dirac operator in the
Grassmann co-ordinate representation. This is achieved by introducing
the ordered symbol for the operator \ct{Ber,Fad,BDW}. A quick way to derive
the necessary results is the following \ct{jw5}. For a single
Grassmann variable $\xi$ the most general form of a differential operator
is

\be
A\, =\, a_0 + a_1 \xi + a_2 \dd{}{\xi} +
          \lh a_3 - a_0 \rh \xi \dd{}{\xi}.
\label{7.9}
\ee

\nit
Consider the action of this operator on an arbitrary function
$f(\xi) = f_0 + f_1 \xi$:

\be
A f(\xi)\, =\, a_0 f_0 + a_2 f_1 + \lh a_1 f_0 + a_3 f_1 \rh \xi.
\label{7.10}
\ee

\nit
The operation of $A$ on $f$ can be represented equivalently in terms of an
integral. First observe that

\be
f(\xi)\, =\, \int d\xi^{\prime}\, \del (\xi^{\prime} - \xi) f(\xi^{\prime})\,
         =\, \int d\xi^{\prime} d\bar{\xi}\, e^{\bar{\xi} (\xi^{\prime} - \xi)}
             f(\xi^{\prime}).
\label{7.11}
\ee

\nit
Here we have introduced the Fourier representation of the anti-commuting
$\del$-function in terms of a conjugate Grassmann variable $\bar{\xi}$.
The ordered symbol $\bar{A}(\xi,\bar{\xi})$ of $A$ is defined by the relation

\be
A f(\xi)\, =\, \int d\xi^{\prime} d\bar{\xi}\,  \bar{A}(\xi,\bar{\xi})
                  e^{\bar{\xi} (\xi^{\prime} - \xi)} f(\xi^{\prime}).
\label{7.12}
\ee

\nit
Our construction shows, that it is obtained by replacing every $\pl/\pl\xi$ in
$A$ by a $\bar{\xi}$:

\be
\bar{A}(\xi,\bar{\xi})\, =\, a_0 + a_1 \xi + a_2 \bar{\xi} +
          \lh a_3 - a_0 \rh \xi \bar{\xi}.
\label{7.13}
\ee

\nit
In the following it is useful to have an expression for the symbol of a product
of operators in terms of the product of their symbols. The relation is given by
the equation

\be
\overline{\left[ AB \right]}(\xi,\bar{\xi})\, =\, \int d\xi^{\prime}
       d\bar{\xi}^{\prime}\, e^{\lh \bar{\xi}^{\prime} - \bar{\xi} \rh
       \lh \xi^{\prime} - \xi \rh }\, \bar{A}(\xi,\bar{\xi}^{\prime})
       \bar{B}(\xi^{\prime}, \bar{\xi}).
\label{7.13.1}
\ee

\nit
It is straightforward to generalize this to a product of an arbitrary number
of operators. In that case a symbolic notation for the integration measure is
used, of the form

\be
\int d^n\xi d^n \bar{\xi}\, \equiv\,
     \int \prod_{\ag = 1}^n\,  d\xi_{\ag} d\bar{\xi}_{\ag}.
\label{7.16.1}
\ee

\nit
Returning to the case of a spinning particle in 4-dimensional space-time we
add two conjugate variables $(\bar{\xi}^1, \bar{\xi}^2)$ and introduce the
symbols for the operators $(\hgm^{\mu}, \hgm_5)$, rescaled by factors
$1/\sqrt{2m}$ for later convenience:

\be
\ba{ll}
\dsp{ \psi^1\, =\, \frac{1}{\sqrt{2m}}\, \lh \xi^1 + \bar{\xi}^1 \rh, } &
\dsp{ \psi^2\, =\, \frac{-i}{\sqrt{2m}}\, \lh \xi^1 - \bar{\xi}^1 \rh, } \\
 & \\
\dsp{ \psi^3\, =\, \frac{1}{\sqrt{2m}}\, \lh \xi^2 + \bar{\xi}^2 \rh, } &
\dsp{ \psi^0\, =\, - i \psi^4\,
               =\, \frac{1}{\sqrt{2m}}\, \lh \xi^2 - \bar{\xi}^2 \rh. }
\ea
\label{7.14}
\ee

\nit
and

\be
\psi_5\, =\, \frac{1}{\sqrt{2m}}\, e^{2\, \bar{\xi} \cdot \xi}\,
               =\, \frac{1}{\sqrt{2m}}\, \lh 2 \xi^1 \bar{\xi}^1 - 1 \rh
                                        \lh 2 \xi^2 \bar{\xi}^2 - 1 \rh .
\label{7.15}
\ee

\nit
Note again the non-standard feature that, whereas the $\psi^{\mu}$ are
Grassmann-odd, the $\psi_5$ introduced here is Grassmann-even (bosonized).

In terms of these quantities, one can rewrite the Dirac equation in momentum
space as

\be
\int d^2\xi^{\prime} d^2\bar{\xi}\, e^{ \bar{\xi} \cdot
     \lh \xi^{\prime} - \xi \rh } \, \left[ p \cdot \psi + m \psi_5 \right]
     (\xi,\bar{\xi})\, \Fg(\xi^{\prime})\, =\, 0.
\label{7.16}
\ee

\nit
As in our representation the chirality equals the Grassmann parity, it
follows that terms with different Grassmann parity in the wave function
are mixed by a non-zero mass.

To obtain the propagator in momentum space we have to invert\footnote{Of
course, this inverse is to be interpreted in the usual sense of distributions
to deal with singularities.} the Grassmann integral operator in
(\ref{7.16}). As a first step, we observe that the product rule (\ref{7.13.1})
implies the usual identity

\be
\int d^2\xi^{\prime} d^2\bar{\xi}^{\prime}\,
 e^{\lh \bar{\xi}^{\prime} - \bar{\xi} \rh \cdot \lh \xi^{\prime} - \xi \rh }\,
 \left[ p \cdot \psi + m \psi_5 \right](\xi,\bar{\xi}^{\prime}) \,
 \left[ p \cdot \psi + m \psi_5 \right](\xi^{\prime},\bar{\xi})\,
 =\,  \frac{p^2 + m^2}{2m}.
\label{7.17}
\ee

\nit
The inverse of the Dirac operator in momentum space can then be written
in an extension of the Schwinger representation as \ct{frad}

\be
\ba{lll}
\left[ S_F\, \hgm_5 \right](p_{\mu};\xi,\bar{\xi}) & = & \dsp{
   \frac{-i}{\sqrt{2m}}\, \int_0^{\infty} dT \int d\sg\,  e^{-\frac{iT}{2m}\,
   \lh p^2 + m^2 - i\ve \rh - \sg\, \lh p \cdot \psi + m \psi_5 \rh }
  }\\
  & & \\
  & = & \dsp{ \sqrt{2m}\;
        \frac{\left[ p \cdot \psi + m \psi_5 \right](\xi,\bar{\xi})}{
        p^2 + m^2 - i\ve }, }
\ea
\label{7.18}
\ee

\nit
where $T$ is the usual Schwinger proper-time parameter, and $\sg$ is a
Grassmann-odd counterpart. Eq.(\ref{7.17}) then implies that

\be
\sqrt{2m}\, \int d^2\xi^{\prime} d^2\bar{\xi}^{\prime}\,
 e^{\lh \bar{\xi}^{\prime} - \bar{\xi} \rh \cdot \lh \xi^{\prime} - \xi \rh }\,
 \left[ p \cdot \psi + m \psi_5 \right](\xi,\bar{\xi}^{\prime}) \,
 \left[ S_F\, \hgm_5 \right] (p_{\mu};\xi^{\prime},\bar{\xi})\, =\, 1.
\label{7.19}
\ee

\nit
In order to get the expression in the co-ordinate representation, we have to
compute the Fourier transform. We also redefine the Grassmann variable,
making it proportional to proper time $T$: $\sg = T \chi$. Now introduce
an integral kernel:

\be
\ba{lll}
K_{\chi}(x - x^{\prime}; \xi, \bar{\xi} |T) & = & \dsp{
     \int \frac{d^4p}{(2\pi)^4}\, e^{i p \cdot (x - x^{\prime})}\,
     e^{ -\frac{iT}{2m}\, \lh p^2 + m^2 - i\ve \rh -
        T \chi \lh p \cdot \psi + m \psi_5 \rh } }\\
  &   & \\
  & = & \dsp{ -i\lh \frac{m}{2\pi T} \rh^{2}\, e^{ \frac{im}{2T}\,
      \lh x - x^{\prime} \rh^2 - \frac{iT}{2}\, (m - i\ve)
       - m \chi \psi \cdot \lh x - x^{\prime} \rh -
       m T \chi \psi_5 }. }
\ea
\label{7.20}
\ee

\nit
Then the following results hold: \nl
$(i)$ The Feynman propagator for a Dirac fermion can be written as

\be
\left[ S_F \hgm_5 \right](x - y; \xi, \bar{\xi})\, =\,
  \frac{-i}{\sqrt{2m}}\, \int_0^{\infty} \frac{dT}{T}\, \int d\chi\,
  K_{\chi}(x - y; \xi, \bar{\xi} |T).
\label{7.21}
\ee

\nit
$(ii)$ Huygens' composition principle holds in the form

\be
\ba{l}
\dsp{ \int d^4x^{\prime} \int d^2\xi^{\prime} d^2\bar{\xi}^{\prime}\,
 e^{\lh \bar{\xi}^{\prime} - \bar{\xi} \rh \cdot \lh \xi^{\prime} - \xi \rh}\,
 K_{\chi}(x-x^{\prime}; \xi, \bar{\xi}^{\prime} |T^{\prime})\,
 K_{\chi}(x^{\prime}-x^{\prime\prime}; \xi^{\prime}, \bar{\xi}
|T^{\prime\prime})
 } \\
   \\
\dsp{ \hspace{5em} =\,
 K_{\chi}(x-x^{\prime\prime}; \xi, \bar{\xi}; |T^{\prime}+T^{\prime\prime}). }
\ea
\label{7.22}
\ee

\nit
$(iii)$ The integral kernel satisfies the boundary condition

\be
K_{\chi}(x-y; \xi, \bar{\xi} |0)\, =\, \del^4(x-y).
\label{7.23}
\ee

\nit
Apart from giving the explicit expression (\ref{7.18}) for the Dirac-Feynman
propagator, eq.(\ref{7.21}) can be used to construct a path-integral
representation of the propagator by iteration of eq.(\ref{7.22}). Indeed,
repeated use of this equation leads to the result

\be
\ba{ll}
\dsp{ K_{\chi}(x - y; \xi, \bar{\xi} |T) } & =\;\;
  \dsp{ \int \prod_{k = 1}^N \left[ d^4 x_k d^2 \xi_k d^2 \bar{\xi}_k \right]
  e^{\frac{1}{2} \sum_{j=1}^N \left[ \lh \bar{\xi}_j - \bar{\xi}_{j-1} \rh
  \cdot \xi_j - \bar{\xi}_j \cdot \lh \xi_{j+1} - \xi_j \rh \right] } }\\
  & \\
  & \dsp{ \times \, e^{+ \frac{1}{2} \lh \bar{\xi}_0 - \bar{\xi}_N \rh \cdot
    \xi_{N+1} - \frac{1}{2} \bar{\xi}_0 \cdot \lh \xi_1 - \xi_{N+1} \rh }\,
    \prod_{s=0}^N\, K_{\chi}(x_{s+1} - x_s; \xi_{s+1}, \bar{\xi}_s | \Del T), }
\ea
\label{7.24}
\ee

\nit
where $x_0 = y$, $x_{N+1} = x$, $\bar{\xi}_0 = \bar{\xi}$, $\xi_{N+1} = \xi$
and $\Del T = T/(N+1)$. Furthermore

\be
\ba{l}
\prod_{s=0}^N\, K_{\chi}(x_{s+1} - x_s; \xi_{s+1}, \bar{\xi}_s | \Del T)\, = \\
  \\
\hspace{4em } =\,
  \dsp{ \left[ \frac{1}{i} \lh \frac{m}{2\pi \Del T} \rh^2 \right]^{N+1}\,
  e^{ \frac{i}{2}\, \sum_{s=0}^N \Del T \left[ m \lh \frac{x_{s+1} - x_s}{
      \Del T} \rh^2 + 2i m \chi \psi_s \cdot \lh \frac{x_{s+1} - x_s}{\Del T}
      \rh + 2i m \chi \psi_s^5  -  m + i\ve \right] } }
\ea
\label{7.25}
\ee

\nit
In agreement with our earlier definitions, the symbols $\psi_k^{\mu}$ here
are a short-hand notation for

\be
\psi_k^1 = \frac{1}{\sqrt{2m}}\, \lh \xi^1_{k+1} + \bar{\xi}^1_k \rh,
\hspace{3em}
\psi_k^2 = \frac{-i}{\sqrt{2m}}\, \lh \xi^1_{k+1} - \bar{\xi}^1_k \rh,
\label{7.26}
\ee

\nit
with similar espressions for $(\psi^3_k, \psi^0_k)$ in terms of $(\xi^2_{k+1},
\bar{\xi}^2_k)$, and the bosonized $\psi^5_k$ given by

\be
\psi_k^5\, =\, \frac{1}{\sqrt{2m}}\, e^{2 \bar{\xi}_k \cdot \xi_{k+1}}.
\label{7.26.1}
\ee

\nit
In the continuum limit ($N \rightarrow \infty, T$ fixed) the free fermion
propagator can now be written as a path integral

\be
\left[ S_F \hgm_5 \right](x - y; \xi, \bar{\xi})\, =\,
  \frac{-i}{\sqrt{2m}}\, \int_0^{\infty} \frac{dT}{T}\, \int d\chi\,
  \cD x(\tau) \cD \xi(\tau) \cD \bar{\xi}(\tau)\,
  e^{i S_{ferm}\left[ x(\tau),\xi(\tau),\bar{\xi}(\tau)\right]},
\label{7.27}
\ee

\nit
where up to boundary terms

\be
\ba{l}
S_{ferm} \left[ x(\tau),\xi(\tau),\bar{\xi}(\tau)\right]\, =\,  \\
   \\
\dsp{ \hspace{4em}
   \int_0^T d\tau\, \left[ \frac{m}{2}\, \dot{x}_{\mu}^2 - \frac{i}{2}\,
    \lh \dot{\bar{\xi}} \cdot \xi - \bar{\xi} \cdot \dot{\xi} \rh +
    i m  \dot{x} \cdot \chi\, \psi(\xi,\bar{\xi}) + i m \chi\,
    \psi_5(\xi,\bar{\xi}) - \frac{m}{2} \right]. }
\ea
\label{7.28}
\ee

\nit
Finally it can be cast into a manifestly Lorentz-invariant form\footnote{
There is a subtlety concerning the Lorentz-invariance of the continuum limit,
as it seems to require that the main contribution to the continuum path
integral comes from paths which are smooth in the sense that for $\Del T
\rightarrow 0$ one has $ \| \xi_{k+1} - \xi_k \| = {\cal O}((\Del
T)^{\frac{1}{2}+p})$ with $p > 0$; from the calculation of explicit examples
directly in the continuum limit this seems to be correct. Some arguments for
a consistent continuum path-integral for simple spin systems, related to the
Duistermaat-Heckman theorem, have been advanced in \ct{ercol}.}
by replacing $(\xi^i(\tau), \bar{\xi}^i(\tau))$ by the vector-like variables
$\psi^{\mu}(\tau)$ defined in eq.(\ref{7.14}):

\be
S_{ferm} \left[ x(\tau),\psi^{\mu}(\tau)\right]\, =\, \frac{m}{2}\,
   \int_0^T d\tau\, \left[ \dot{x}_{\mu}^2 + i \psi \cdot \dot{\psi}
    + 2 i \dot{x} \cdot \chi\, \psi + 2 i \chi\, \psi_5 - 1 \right].
\label{7.29}
\ee

\nit
This expression resembles closely the one usually encountered in the
literature \ct{brink1,brink2}, \ct{BerMar}-\ct{jw4}, which is based on
the realization of a (gauge-fixed) local world line supersymmetry.
However, in contrast to the standard approach the present construction
uses a Grassmann-{\em even} $\psi_5$, and as a result inclusion of the
mass-term violates explicit supersymmetry at the classical level, even
though it is realized on the operator level in the quantum theory, as
shown by eq.(\ref{7.17}). In fact, for non-vanishing $m$ the
pseudo-classical action $S_{ferm}$ does not even have a well-defined
Grassmann parity; eq.(\ref{7.4.1}) and the discussion following it makes
clear that this is a direct consequence of the non-conservation of
chirality for massive fermions.

It is shown below, that the discrepancy between the result (\ref{7.28})
and the manifestly supersymmetric spinning particle model has its origin in
a doubling of the number of degrees of freedom in the supersymmetric case,
which is regularly overlooked. More precisely, if one preserves supersymmetry
of the classical action by taking $\psi_5$ to be Grassmann-odd, and if the
classical algebra of Poisson-Dirac brackets is mapped in a straightforward
way to the quantum (anti-)commutation relations, then one does not obtain
the representation of the quantum operator $\hat{\psi}_5$ by the 4-dimensional
Dirac matrix $\gam_5 = -i/4!\, \ve_{\mu\nu\kg\lb} \gam^{\mu} \gam^{\nu}
\gam^{\kg} \gam^{\lb}$. Rather, a Grassmann-odd (fermionic) representation of
the operator $\hat{\psi}_5$ is to be used, anti-commuting with the other
$\hat{\psi}^{\mu}$. Then the equivalent matrix representation of this
operator is a $\gam_5$ taken from the Dirac algebra in six dimensions.
This implies a doubling of the number of degrees of freedom of an irreducible
spinor.

On the other hand, for free massless particles manifest supersymmetry is not
violated even in the present theory; this is because chirality is conserved,
and so only terms of even Grassmann parity occur in the action. One therefore
sees that the apparent violation of manifest world line supersymmetry is a
result of mass generation, and has the same dynamical origin.

\section{Worldline supersymmetry}{\label{S.8}}

In this section we turn to the point-particle model with full classical
super-reparametrization invariance on the world line, to compare it
with our treatment of Dirac fermions in sect.(\ref{S.7}). To realize
complete off-shell supersymmetry\footnote{The term {\em off shell}
implies that the supersymmetry algebra is realized without using
dynamical constraints like the equations of motion.} it is necessary
to introduce three different types of super-multiplets (superfields):

\begin{enumerate}

\item{ The gauge multiplet $(e, \chi)$ consisting of the einbein $e$ and its
superpartner $\chi$, also refered to as gravitino; under local worldline
supersymmetry with parameter $\ve$ the multiplet transforms as

\be
\del e\, =\, -2i \ve \chi, \hspace{3em}
\del \chi\, =\, \frac{d\ve}{d\tau}.
\label{8.1}
\ee
}

\item{ Matter multiplets $(x^{\mu}, \psi^{\mu})$ describing the position
and spin co-ordinates of the particle. The transformation rules are

\be
\del x^{\mu}\, =\, -i \ve \psi^{\mu}, \hspace{3em}
\del \psi^{\mu}\, =\, \ve \frac{1}{e}\, \frac{Dx^{\mu}}{D\tau},
\label{8.2}
\ee

\nit
where the super-covariant derivative is constructed with the gravitino
as the connection:

\be
\frac{Dx^{\mu}}{D\tau}\, =\, \frac{dx^{\mu}}{d\tau}\, +\, i \chi \psi^{\mu}.
\label{8.3}
\ee
}

\item{ One or more fermionic multiplets $(\eta, f)$ with Grassmann-odd $\eta$
and even $f$; it is used in the following as an auxiliary multiplet and its
transformation properties under local supersymmetry are

\be
\del \eta\, =\, \ve f, \hspace{3em}
\del f\, =\, -i \ve \frac{1}{e}\, \frac{D\eta}{D\tau}.
\label{8.4}
\ee

\nit
The super-covariant derivative is formed as before:

\be
\frac{D\eta}{D\tau}\, =\, \frac{d\eta}{d\tau}\, -\, \chi f.
\label{8.5}
\ee
}

\end{enumerate}

\nit
It may be checked that in all cases the commutator of two supersymmetry
transformations gives a local translation (reparametrization) with
parameter $a = (2 i/e) \ve_1 \ve_2$, plus a local supersymmetry transformation
with parameter $\ve^{\prime} = -(2 i/e) \ve_1 \ve_2 \chi$.

The minimal free particle action invariant under these one-dimensional local
supersymmetry transformations is

\be
S_{susy} =  \frac{m}{2}\, \int d\tau\, \lh
            \frac{1}{e}\, \dot{x}_{\mu}^2 + i \psi \cdot \dot{\psi} +
            \frac{2i}{e}\, \chi \psi \cdot \dot{x} + i \eta \dot{\eta} +
            2i \chi \eta + e f^2 - 2 e f \rh
\label{8.6}
\ee

\nit
Obviously, the bosonic variable $f$ appearing quadratically without derivatives
does not represent a dynamical degree of freedom and may be eliminated using
its algebraic equation of motion $f = 1$ (which is equivalent to completing
the square). Then the classical action becomes

\be
S_{susy}\,  =\, \frac{m}{2}\, \int d\tau\, \lh
               \frac{1}{e}\, \dot{x}_{\mu}^2 + i \psi \cdot \dot{\psi} +
               + i \eta \dot{\eta} + \frac{2i}{e}\, \chi \psi \cdot \dot{x}
               + 2 i \chi \eta - e \rh.
\label{8.7}
\ee

\nit
In addition local super-reparametrization invariance may be used to fix the
gauge multiplet $(e,\chi)$ to constant values. In particular, if the proper
time $\tau$ is rescaled by $e$, so as to be measured in the same units as the
laboratory time (effectively setting $e = 1$), and the constant value of
$\chi$ is rescaled by the same amount, then comparison with the fermionic
action (\ref{7.28}) shows that $S_{ferm}$ is formally related to $S_{susy}$
by $\dot{\eta} \rightarrow 0$, $\eta \rightarrow \psi_5$. However, this last
relation is frustrated by the mismatch in the Grassmann parity of the
two quantities; moreover, contrary to $f$ the variable $\eta$ is a dynamical
degree of freedom, associated with an additional two-valued parameter labeling
the wave-functions of the particle. As it couples to the gravitino, and
therefore appears in the first-class constraints of the theory, its dynamics
cannot be taken into account by simply equating it to a constant in the action.
At this point we observe however, that one could add more fermionic multiplets
$(\eta_k, f_k)$ whose auxiliary scalars have vanishing classical value:

\be
\Del S\, =\, \frac{m}{2}\, \int d \tau\, \lh i \eta_k \dot{\eta}_k +
             e f_k^2 \rh.
\label{8.7.1}
\ee

\nit
Eliminating the $f_k$ by their algebraic Euler-Lagrange equation $f_k = 0$,
there remain only free fermionic degrees of freedom $\eta_k$ which decouple
from the dynamics because they do not interact with the other physical or the
gauge degrees of freedom. In fact these additional fermions define a $d = 1$
topological field theory, in the sense that their action is reparametrization
invariant without involving the metric (represented by the einbein $e$). It
is also easy to see, that they do not contribute to the hamiltonian of the
theory. \vs

\nit
To construct the canonical quantum theory corresponding to the supersymmetric
action $S_{susy}$, we employ the BRST procedure used before. Besides the
reparametrization ghosts $(b,c)$ and the lagrange multiplier $\lb$ we
introduce commuting supersymmetry ghosts $(\ag,\bg)$ and an anti-commuting
multiplier $s$. The nilpotent BRST variations of the full set of variables
read:

\begin{enumerate}

\item{For the gauge multiplet:

\be
\dL e = \frac{d(ce)}{d\tau} + 2 \ag \chi, \hspace{3em}
\dL \chi = \frac{d(c\chi)}{d\tau} + i \dot{\ag}.
\label{8.8}
\ee
}

\item{For the matter multiplets:

\be
\dL x^{\mu} = c \dot{x}^{\mu} + \ag \psi^{\mu}, \hspace{3em}
\dL \psi^{\mu} = c \dot{\psi}^{\mu} + \frac{i\ag}{e}\, \frac{Dx^{\mu}}{D\tau}.
\label{8.9}
\ee
}

\item{For the fermionic multiplet:

\be
\dL \eta = c \dot{\eta} + i \ag f, \hspace{3em}
\dL f = c \dot{f} + \frac{\ag}{e}\, \frac{D\eta}{D\tau}.
\label{8.10}
\ee
}

\item{For the ghost variables:

\be
\ba{ll}
\dsp{ \dL c = c \dot{c} - \frac{i \ag^2}{e}, } &
\dsp{ \dL \ag = c \dot{\ag} + \frac{\ag^2}{e}\, \chi, } \\
  & \\
\dsp{ \dL b = - i \lb, } & \dsp{ \dL \bg = s, } \\
  & \\
\dsp{ \dL \lb = 0, } & \dsp{ \dL s = 0. }
\ea
\label{8.11}
\ee
}

\end{enumerate}

\nit
In the classical theory defined by $S_{susy}$ we wish to impose the gauge
conditions

\be
\dot{e}\, =\, 0, \hspace{3em} \dot{\chi}\, =\, 0.
\label{8.13}
\ee

\nit
Two write down a relatively simple BRST-invariant gauge-fixed action,
it is convenient to make the redefinitions:

\be
ec\, \rightarrow\, c, \hspace{3em}
\ag - i c \chi\, \rightarrow\, \ag.
\label{8.12}
\ee

\nit
Then the BRST variations are covariantized and simplified; for example

\be
\ba{ll}
\dL \chi = i \dot{\ag}, & \dL c = - i \ag^2, \hspace{3em} \dL \ag = 0, \\
 & \\
\dsp{ \dL x^{\mu} = \frac{c}{e}\, \frac{Dx^{\mu}}{D\tau} + \ag \psi^{\mu}, } &
\dsp{ \dL \psi^{\mu} = \frac{c}{e}\, \frac{D\psi^{\mu}}{D\tau} +
      \frac{i\ag}{e}\, \frac{Dx^{\mu}}{D\tau}, } \\
 & \\
\dsp{ \dL \eta = \frac{c}{e}\, \frac{D\eta}{D\tau} + i \ag f, } &
\dsp{ \dL f = \frac{c}{e}\, \frac{Df}{D\tau} + \frac{\ag}{e}\,
      \frac{D\eta}{D\tau}. }
\ea
\label{8.13.1}
\ee

\nit
In terms of these new variables the BRST-invariant gauge-fixed action reads

\be
S_{gf}\, =\, S_{susy}\, +\, \int_0^T d\tau \lh \lb \dot{e} + i s \dot{\chi}
         + i \dot{b} \dot{c} + \dot{\bg} \dot{\ag} + 2 i \ag \dot{b} \chi \rh,
\label{8.14}
\ee

\nit
where we take $S_{susy}$ as in eq.(\ref{8.7}), in which $f = 1$
has been inserted. This is consistent with the BRST variations (\ref{8.13.1})
and their nilpotency, provided the equation of motion $\dot{\eta} = \chi$ is
used.

The canonical momenta for the physical and gauge variables in this theory are

\be
\ba{ll}
\dsp{ p^{\mu} = \frac{m}{e}\, \frac{Dx^{\mu}}{D\tau} , } & \\
 & \\
\dsp{ \pi^{\mu} = - \frac{im}{2}\, \psi^{\mu}, } &
\dsp{ \pi_{\eta} = - \frac{im}{2}\, \eta. } \\
 & \\
\dsp{ p_e = \lb, }  & \dsp{ \pi_{\chi} = - i s, }
\ea
\label{8.15}
\ee

\nit
This is to be supplemented with the ghost momenta

\be
\ba{ll}
\dsp{ \pi_c = - i \dot{b}, } &
\dsp{ \pi_b = i \lh \dot{c} + 2 \ag \chi \rh, } \\
 & \\
\dsp{ p_{\ag} = \dot{\bg}, } &
\dsp{ p_{\bg} = \dot{\ag}. }
\ea
\label{8.16}
\ee

\nit
A minor complication is the appearance of the second-class constraints
for the fermionic momenta $(\pi^{\mu}, \pi_{\eta})$. These are resolved
in the standard way by Dirac's procedure, and one ends up with the gauge-fixed,
unconstrained hamiltonian

\be
H_{gf}\, =\, \frac{e}{2m}\, \lh p_{\mu}^2 + m^2 \rh\,  -\,
             i \chi \lh \psi \cdot p + m \eta - 2 i \ag \pi_c \rh
              -\, i \pi_b \pi_c\, +\, p_{\ag} p_{\bg},
\label{8.17}
\ee

\nit
plus the following Dirac-Poisson bracket for functions $(F, G)$ on the
unconstrained phase-space spanned by $(x^{\mu}, p_{\mu}; \psi^{\mu};
\eta; e, p_e; \chi, \pi_{\chi}; c, \pi_c; b, \pi_b; \ag, p_{\ag}; \bg,
p_{\bg})$:

\be
\ba{lll}
\left\{F, G \right\} & = & \dsp{ \dd{F}{x^{\mu}} \dd{G}{p_{\mu}}\, -\,
     \dd{F}{p_{\mu}} \dd{G}{x^{\mu}}\, +\, \dd{F}{e} \dd{G}{p_e}\, -\,
     \dd{F}{p_e} \dd{G}{e} } \\
 & & \\
 & & \dsp{
        +\, \dd{F}{\ag} \dd{G}{p_{\ag}}\, -\, \dd{F}{p_{\ag}} \dd{G}{\ag}\, +\,
        \dd{F}{\bg} \dd{G}{p_{\bg}}\, -\, \dd{F}{p_{\bg}} \dd{G}{\bg}\, } \\
 & & \\
 & & \dsp{ +\, (-1)^{a_F} \lh \frac{i}{m}\, \dd{F}{\psi^{\mu}}
     \dd{G}{\psi_{\mu}} + \frac{i}{m}\, \dd{F}{\eta} \dd{G}{\eta} +
     \dd{F}{\chi} \dd{G}{\pi_{\chi}} + \dd{F}{\pi_{\chi}} \dd{G}{\chi} \rh } \\
 & & \\
 & & \dsp{ +\, (-1)^{a_F} \lh \dd{F}{c} \dd{G}{\pi_c} + \dd{F}{\pi_c} \dd{G}{c}
        \, +\, \dd{F}{b} \dd{G}{\pi_b} + \dd{F}{\pi_b} \dd{G}{b} \rh. }
\ea
\label{8.18}
\ee

\nit
The equations of motion then take the canonical form

\be
\frac{dF}{d\tau}\, =\, \left\{ F, H_{gf} \right\}.
\label{8.19}
\ee

\nit
Similarly, the BRST variation of a phase-space function $F$ can now be
obtained from

\be
\dL F\, =\,  (-1)^{a_F} \left\{ F, \Og \right\},
\label{8.20}
\ee

\nit
where $\Og$ is the nilpotent, conserved BRST charge

\be
\ba{c}
\dsp{ \Og\, =\, \frac{c}{2m}\, \lh p^2_{\mu} + m^2 \rh\, +\,
          \ag \lh p \cdot \psi + m \eta \rh\, - i \pi_b p_e\, +\,
          i p_{\bg} \pi_{\chi}\, -\, i \ag^2 \pi_c, } \\
   \\
\dsp{ \left\{ \Og, H_{gf} \right\}\, =\, 0, \hspace{3em}
       \left\{ \Og, \Og \right\}\, =\, 0. }
\ea
\label{8.21}
\ee

\nit
Like for the scalar particle, the full hamiltonian $H_{gf}$ is BRST-exact:

\be
H_{gf}\, =\, \left\{ - e \pi_c + i \chi p_{\ag}, \Og \right\}.
\label{8.22}
\ee

\nit
In the BRST cohomology the hamiltonian is therefore equivalent to zero and the
physical particle motions are on the mass-shell.

\section{Quantum supersymmetry}{\label{S.9}}

The BRST-invariant classical action for the supersymmetric theory can now
be taken as the starting point for the construction of a canonical
quantum theory for a free point particle. Later this quantum theory is then
used to construct a propagator, and it is shown that it can be expressed in
path-integral form using precisely the classical action $S_{gf}$. Thus the
correspondence between the classical and quantum theory is established
in both directions. From the canonical formulation it will then be entirely
clear, that for $m \neq 0$ this theory describes a degenerate pair of
Dirac fermions, and that this is completely independend of the BRST
quantization procedure; the origin of the doubling of the degrees of freedom
can in fact be traced to the fermionic dynamical variable $\eta$.

The first step in defining the quantum theory is to postulate a set of
(anti) commutation relations in correspondence with the classical Dirac-Poisson
brackets (\ref{8.18}). In addition to the operator (anti) commutators
(\ref{5.1}), we introduce the operator algebra

\be
\ba{rlrl}
\left\{ \hps_{\mu}, \hps_{\nu} \right\} & \dsp{
                    =\; \frac{1}{m} \eta_{\mu\nu}, } &
\left\{ \hps_6, \hps_6 \right\} & \dsp{ =\; \frac{1}{m}, } \\
 & & & \\
\left\{ \hch, \hat{\pi}_{\chi} \right\} & =\; - i, & &  \\
 & & & \\
\left[ \hag, \hp_{\ag} \right] & =\; i, &
\left[ \hbg, \hp_{\bg} \right] & =\; i.
\ea
\label{9.1}
\ee

\nit
The notation $\hps_6$ (rather than the more usual $\hps_5$) has been introduced
for the operator corresponding to $\eta$, to emphasize the Clifford algebra
structure of the fermion anti-commutators, but also to avoid thinking of this
operator as the product of the other $\hps^{\mu}$ as in the case of the
Dirac fermion in sect.(\ref{S.7}). The above commutation relations can
be satisfied by choosing the following {\em linear} representation of the
operators: the $\hps^{\mu}$ are realized in terms of two Grassmann variables
$(\xi^1, \xi^2)$ and their derivatives, as in (\ref{7.3.1}) by

\be
\hps_{\mu}\, =\, \frac{1}{\sqrt{2m}}\, \hgm_{\mu}(\xi^1,\xi^2).
\label{9.1.1}
\ee

\nit
As before, the operators $\hps_0$ are hermitean only w.r.t.\ to the
physical indefinite-metric inner-product involving Pauli-conjugate spinors.

Furthermore we introduce Grassmann variables $\xi^3$ and $\chi$,
as well as ordinary real variables $(\ag,\bg)$ and define

\be
\ba{llll}
\hps_6 & \dsp{ =\; \frac{1}{\sqrt{2m}}\, \lh \xi^3 + \dd{}{\xi^3} \rh, } &
\hat{\pi}_{\chi} & \dsp{ =\; - i \dd{}{\chi}, } \\
 & & & \\
\hp_{\ag} & \dsp{ =\; - i \dd{}{\ag}, } &
\hp_{\bg} & \dsp{ =\; - i \dd{}{\bg}. }
\ea
\label{9.2}
\ee

\nit
Note that we have a fermionic (Grassmann-odd) representation of $\hps_6$.
As a result, having at our disposition the variable $\xi^3$, one can
define at no expense the additional operator

\be
\hps_5\, =\, - \frac{i}{\sqrt{2m}}\, \lh \xi^3 - \dd{}{\xi^3} \rh,
\label{9.2.1}
\ee

\nit
which anti-commutes with the other $\hps_{M}$. This operator does not
appear in the hamiltonian of the theory, which we define in correspondence
with the classical hamiltonian (\ref{8.17}):

\be
\ba{lll}
\hH_{gf} & = & \dsp{ \frac{e}{2m}\, \lh -\Box  + m^2 \rh\,  -\,
         i \chi \lh -i \hps \cdot \pl + m \hps_6 \rh\,
         +\, 2 i \ag \chi \dd{}{c}\, +\, i \dd{^2}{b \pl c}\, -\,
         \dd{^2}{\ag \pl \bg} } \\
 & & \\
 & \equiv & \dsp{ e \hH_0\, - \, i \chi \hQ_+\, +\, 2 i \ag \chi \dd{}{c}\, +\,
             i \dd{^2}{b \pl c}\, -\, \dd{^2}{\ag \pl \bg}, }
\ea
\label{9.3}
\ee

\nit
and a nilpotent BRST operator

\be
\ba{lll}
\hOg & = & \dsp{  \frac{c}{2m}\, \lh - \Box + m^2 \rh\, +\,
     \ag \lh -i \hps \cdot \pl + m \hps_6 \rh\, +\,
     i \dd{^2}{e \pl b}\, -\, i \dd{^2}{\bg \pl \chi}\, -\,
     \ag^2 \dd{}{c} } \\
 & & \\
 & = & \dsp{ c \hH_0\, +\, \ag \hQ_+\,  +\, i \dd{^2}{e \pl b}\, -\,
             i \dd{^2}{\bg \pl \chi}\, -\, \ag^2 \dd{}{c}. }
\ea
\label{9.4}
\ee

\nit
The short-hand notation in the second line of these equations has been
introduced with a view to the supersymmetry algebra

\be
\hQ_+^2\, =\, \hH_0.
\label{9.5}
\ee

\nit
At this point a remark of caution is appropriate: for $m \neq 0$ the operator
$\hH_0$ has zero-modes only in Minkowski space; but for lorentzian metrics
the operator $\hQ_+$ (essentially the Dirac operator) is not hermitean w.r.t.\
the positive-definite inner-product $\Psi^{\dagger} \Psi$ (only w.r.t.\ the
Lorentz-invariant indefinite inner product $\overline{\Psi} \Psi$). Therefore
the zero-modes of $\hH_0$ (the physical states) are not necessarily zero-modes
of $\hQ_+$: one also finds back the negative-energy states associated with the
vanishing of $\hQ_+^{\dagger}$. Moreover, there is a similar relation for the
Lorentz-invariant operator $\hQ_-$ obtained by replacing $m \rightarrow -m$:

\be
\hQ_-^2\, =\, \hH_0, \hspace{3em}
\hQ_-\, =\, \lh - i \hps \cdot \pl - m \hps_6 \rh.
\label{9.6}
\ee

\nit
The resolution of these difficulties lies of course in the full
quantum-field theoretical treatment; it is of interest to see how the
BRST procedure is implemented in this context, and as a first step we
construct in this paper the free propagator, in sect.(\ref{S.10}).

For our present purpose it is however sufficient to note, that the
BRST-cohomology can still be used to characterize the physical states, in
the following way: since the hamiltonian $H_{gf}$ is a BRST-exact operator:

\be
H_{gf}\, =\, \left\{ e \dd{}{c} - i \chi \dd{}{\ag}, \hOg \right\},
\label{9.7}
\ee

\nit
one can pick a state from each equivalence class of solutions of the
BRST condition

\be
\hOg \Psi_{phys}\, =\, 0,
\label{9.8}
\ee

\nit
by requiring the subsidiary condition

\be
\lh e \dd{}{c} - i \chi \dd{}{\ag} \rh\, \Psi_{phys}\, =\, 0,
\label{9.9}
\ee

\nit
for {\em arbitrary} values of $e$ and $\chi$, as suggested by
super-reparametrization invariance. This amounts actually to two
subsidiary conditions:

\be
\dd{}{c}\, \Psi_{phys}\, =\, 0, \hspace{3em}
\dd{}{\ag}\, \Psi_{phys}\, =\, 0.
\label{9.10}
\ee

\nit
One may think of these conditions as defining the ghost vacuum. It now
follows automatically from (\ref{9.7}) that

\be
H_{gf}\, \Psi_{phys}\, =\, 0.
\label{9.11}
\ee

\nit
Combined with the BRST invariance of the physical states expressed by
(\ref{9.8}), this gives the physical states precisely as solutions of
the Dirac (and consequently Klein-Gordon) equation:

\be
\lh - i \hps \cdot \pl + m \hps_6 \rh\, \Psi_{phys}\, =\, 0.
\label{9.12}
\ee

\nit
The wave functions $\Psi_{phys}$ can be decomposed as

\be
\Psi_{phys}(\xi^1,\xi^2,\xi^3)\, =\, \Phi_1(\xi^1,\xi^2)\, +\,
                                     \xi^3 \Phi_2(\xi^1,\xi^2),
\label{9.13}
\ee

\nit
where each of the terms $\Phi_{1,2}(\xi^1,\xi^2)$ is a 4-component spinor of
the type (\ref{7.4}). Therefore the physical states have eight spinor
components rather than four. Working out the action of the Dirac operator on
these wave functions, eq.(\ref{9.12}), it can be written explicitly in matrix
notation, in terms of ordinary four-dimensional Dirac matrices, as

\be
\lh \ba{cc}  i \gam_5 \gam \cdot p & m \\
             m & - i \gam_5 \gam \cdot p \\
    \ea \rh\, \left[ \ba{c} \Fg_1 \\
                            \Fg_2 \\
                     \ea \right]\, =\,
\lh \sum_{\mu = 0}^3\, \Gam_{\mu} p^{\mu}\, +\, m \Gam_6\, \rh\, \Psi\, =\, 0,
\label{9.14}
\ee

\nit
where the $\Gam_M$ are an 8-dimensional representation of the Dirac matrices
in 6-dimensional space-time. Thus the supersymmetric spinning particle can
be thought of as a reduction of a 6-dimensional massless spinor to 4
space-time dimensions, by compactification on a circle in the 6th dimension
$(p^6 = m)$ and trivial in $x^5$ $(p^5 = 0)$. It follows, that the states of
this theory represent a degenerate pair of 4-dimensional massive fermions.

\section{Supersymmetric propagator}{\label{S.10}}

In the co-ordinate representation, the quantum states of the supersymmetric
particle in the full extended state space (including the ghost degrees of
freedom) are represented by wave functions depending on the variables
$ Z = (x^{\mu}, \xi^k, e, \chi, c, b, \ag, \bg)$, with $\mu = 0,...,3$ and
$k = 1,2,3$. To obtain the second quantized propagator for this theory, we
first construct the evolution operator associated with the hamiltonian
$\hH_{gf}$, eq.(\ref{9.3}).

It is actually convenient to do this in two steps: first we construct the
expression for the matrix elements of $\hK(T) = e^{-iT\hH_{gf}}$ restricted to
the space of gauge- and ghost degrees of freedom $z = (e, \chi; c, b; \ag,
\bg)$, leaving the `matter' content $(x^{\mu}, \xi^k)$ unspecified; only then
we specify the precise model for the physical degrees of freedom. The
advantage of this procedure is, that the results are easily applied to other
models, for example particles interacting with background fields.

Our starting point is the general expression (\ref{9.3}) for $\hH_{gf}$:

\[
\hH_{gf}\, =\, e \hH_0\, - \, i \chi \hQ_+\, +\, 2 i \ag \chi \dd{}{c}\, +\,
             i \dd{^2}{b \pl c}\, -\, \dd{^2}{\ag \pl \bg}, \]

\nit
where $\hH_0 = \hQ_+^2$. For any such $\hH_0$ and $\hQ_+$ not depending on the
gauge- and ghost degrees of freedom the following results hold:

\begin{enumerate}

\item{ The equation

\be
i \dd{}{T}\, \hK(z,\pz|T)\, =\, \hH_{gf }\, \hK(z,\pz|T),
\label{10.1}
\ee

\nit
with the initial condition

\be
\hK (z,\pz|0)\, =\, \del(e - e^{\prime}) \del(\chi - \chi^{\prime})
  \del(c - c^{\prime}) \del(b - b^{\prime}) \del(\ag -\ag^{\prime})
  \del(\bg - \bg^{\prime})\, \hI,
\label{10.2}
\ee

\nit
has the solution

\be
\ba{ll}
\hK(z,\pz|T)\:\: = & \dsp{ \frac{1}{2\pi}\, \del(e - e^{\prime})
    \del(\chi - \chi^{\prime})\, e^{\frac{i}{T}\, (\ag - \ag^{\prime})
    (\bg - \bg^{\prime}) - (\ag + \ag^{\prime})  (b - b^{\prime}) \chi
    - \frac{1}{T}\, (b - b^{\prime}) (c - c^{\prime})} } \\
 & \\
 & \dsp{ \times\,  e^{- i T \lh e \hH_0 - i \chi \hQ_+ \rh}. }
\ea
\label{10.3}
\ee
}

\item{ This solution satisfies the composition rule

\be
\int d\pz\, \hK(z,\pz|T^{\prime})\,
         \hK(\pz,z^{\prime\prime}|T^{\prime\prime})\, =\,
         \hK(z,z^{\prime\prime}|T^{\prime} + T^{\prime\prime}).
\label{10.4}
\ee
}

\end{enumerate}

\nit
We observe that the gauge $(e,\chi) = const.$ is manifestly realized
in the expression (\ref{10.3}). Let us now first derive an operator
expression for the propagator in the general case (before specifying
$\hH_0$ and $\hQ_+$):

\be
\hDel_{gf}(z,z^{\prime})\, =\,
          \frac{ie}{2m}\, \int_0^{\infty} dT \hK(z,z^{\prime}|T),
\label{10.4.1}
\ee

\nit
which is a solution of the generalized Klein-Gordon-Dirac equation

\be
 \hH_{gf}\, \hDel_{gf}(z,z^{\prime})\, =\, \frac{e}{2m}\, \del(z - \pz) \hI.
\label{10.4.2}
\ee

\nit
To get an explicit expression for $\hDel_{gf}$ we use the following
intermediate results:

\be
e^{\frac{i}{T}\, \lh \ag - \ag^{\prime} \rh \lh \bg - \bg^{\prime} \rh }\,
 =\, \frac{T}{2\pi}\, \int dp_{\ag} dp_{\bg}\,
 e^{i p_{\ag} \lh \ag - \ag^{\prime} \rh + i p_{\bg} \lh \bg - \bg^{\prime} \rh
 - i T p_{\ag} p_{\bg} },
\label{10.4.3}
\ee

\nit
and

\be
e^{- \frac{1}{T}\, \lh b - b^{\prime} \rh \lh c - c^{\prime} \rh }\,
 =\, \frac{1}{T}\, \int d\pi_b d\pi_c\,
 e^{\lh b - b^{\prime} \rh \pi_b + \pi_c \lh c - c^{\prime} \rh
 - T \pi_b \pi_c}.
\label{10.4.4}
\ee

\nit
Note that the factors of $T$ in front of these ghost kernels cancel in
the expression for $\hK(T)$, as expected from supersymmetry. It is now
straightforward to obtain the operator expression for the propagator in the
ghost-momentum space:

\be \ba{l}
\dsp{ \hDel_{gf}(z,z^{\prime})\, =\, \frac{e}{8 \pi^2 m}\,
 \del(e - e^{\prime}) \del(\chi - \chi^{\prime})\,
 e^{- (\ag + \ag^{\prime}) (b - b^{\prime}) \chi}\, \times } \\
 \\
 \dsp{ \hspace{1em}
 \int dp_{\ag} dp_{\bg} \int d\pi_b d\pi_c\, e^{i p_{\ag} (\ag - \ag^{\prime})
 + i p_{\bg} (\bg - \bg^{\prime}) + (b - b^{\prime}) \pi_b +
 \pi_c (c - c^{\prime})}\, \left[e \hH_0 - i \chi \hQ_+ + p_{\ag} p_{\bg}
 - i \pi_b \pi_c \right]^{-1}. }
\ea
\label{10.4.5}
\ee

\nit
The matrix elements of the inverse operator inside the square brackets
now have to be computed. Again, we first consider the evolution operator.
For the free particle we use the elementary result

\be
\langle x | e^{- iT \lh  e \hH_0 - i \chi \hQ_+ \rh} | x^{\prime} \rangle\,
  =\, \int \frac{d^4p}{(2\pi)^2}\, e^{i p \cdot \lh x - x^{\prime} \rh}\,
  e^{- \frac{ieT}{2m}\, \lh p^2_{\mu} + m^2 \rh - T \chi \lh \hps \cdot p +
  m \hps_6 \rh }.
\label{10.5}
\ee

\nit
In addition we replace the fermionic operators $(\hps^{\mu}, \hps_6)$ by
their symbols as in eq.(\ref{7.14}), supplemented by

\be
\eta\, =\, \frac{1}{\sqrt{2m}}\, \lh \xi^3\, +\, \bar{\xi}^3 \rh,
\label{10.6}
\ee

\nit
where we have returned to the original notation for the additional
fermionic degree of freedom. Similarly, for later convenience, we also
introduce the symbol of the presently redundant operator $\hps_5$,
denoting it by $\eta_1$:

\be
\eta_1\, =\, - \frac{i}{\sqrt{2m}}\, \lh \xi^3\, -\, \bar{\xi}^3 \rh.
\label{10.6.1}
\ee

\nit
Carrying out the integration over momentum $p_{\mu}$ then gives the
expression for the matrix element $K(Z,\pZ|T)$:

\be
\ba{ll}
K_{\ve}(Z,\pZ|T)\:\: = & \dsp{ \frac{1}{2\pi}\, \del(e - e^{\prime})
    \del(\chi - \chi^{\prime})\, e^{\frac{i}{T}\, (\ag - \ag^{\prime})
    (\bg - \bg^{\prime}) - (\ag + \ag^{\prime})  (b - b^{\prime}) \chi
    - \frac{1}{T}\, (b - b^{\prime}) (c - c^{\prime})} } \\
 & \\
 & \dsp{ \times\, \left[ -i \lh \frac{m}{2\pi e T} \rh^2\,
   e^{\frac{im}{2eT}\, \lh x - x^{\prime} \rh^2 - \frac{ieT}{2}\,
   \lh m - i\ve \rh - \frac{m}{e}\, \chi \psi \cdot \lh x - x^{\prime} \rh
   - m T \chi \eta } \right]. }
\ea
\label{10.7}
\ee

\nit
Note that after rescaling $T$ and $\chi$ by a factor $e$ (or equivalently,
taking $e = 1$), the expression in brackets involving the physical
degrees of freedom is almost identical with the expression (\ref{7.20}) for
the kernel $K_{\chi}(T)$ of the single Dirac fermion, except for the
replacement of the Grassmann-even $\psi_5(\xi^i,\bar{\xi}^i)$, $(i=1,2)$,
by the Grassmann-odd $\eta(\xi^3,\bar{\xi}^3)$.

The expression for the propagator $\Del_{gf}(Z,\pZ)$ in the full ghost-extended
state space is similarly obtained by taking the matrix element of
$\hDel(z,z^{\prime})$; this amounts to making the replacement

\be
\ba{l}
\dsp{ \frac{e}{2m}\, \left[e \hH_0 - i \chi \hQ_+ + p_{\ag} p_{\bg}
  - i \pi_b \pi_c \right]^{-1}\, =\,
  \frac{ie}{2m}\, \int_0^{\infty} dT\, e^{-iT
  \left[e \hH_0 - i \chi \hQ_+ + p_{\ag} p_{\bg} - i \pi_b \pi_c \right]} } \\
  \\
 \hspace{2em}  \rightarrow \dsp{
  \int \frac{d^4p}{(2\pi)^2}\, \frac{e^{i p \cdot (x - x^{\prime})}}{
  p^2 + m^2 - i \ve + \frac{2m}{e}\, \left[ -i \chi (p \cdot \psi + m \eta )
  - i \pi_b \pi_c + p_{\ag} p_{\bg} \right]} }
\ea
\label{10.7.1}
\ee

\nit
in the r.h.s.\ of eq.(\ref{10.4.5}), and interpreting $(\psi^{\mu}, \eta)$ as
the above symbols.

The construction of the path-integral formula for the propagator now
repeats the steps for the Dirac fermion, with the difference that instead of
the symbol of the physical part depending on two canonical pairs of Grassmann
variables $(\xi^k,\bar{\xi}^k)$, there are now three such pairs. Then of
course there is also the difference that we have included here additional
ghost degrees of freedom. However, these do not cause any difficulties.
As a result we obtain

\be
\ba{ll}
\dsp{ K_{\ve}(Z,\pZ |T) } & =\;\; \dsp{ \int \left[ \prod_{k = 1}^N dZ_k
  \right]\; e^{\frac{1}{2} \sum_{j=1}^N \left[ \lh \bar{\xi}_j -
  \bar{\xi}_{j-1} \rh \cdot \xi_j - \bar{\xi}_j \cdot \lh \xi_{j+1} -
  \xi_j \rh \right] } }\\
  & \\
  & \dsp{ \times \, e^{+ \frac{1}{2} \lh \bar{\xi}_0 - \bar{\xi}_N \rh \cdot
    \xi_{N+1} - \frac{1}{2} \bar{\xi}_0 \cdot \lh \xi_1 - \xi_{N+1} \rh }\,
    \prod_{s=0}^N\, K_{\ve}(Z_{s+1}, Z_s | \Del T), }
\ea
\label{10.8}
\ee

\nit
where the exponent involves 3 types of $(\xi^k,\bar{\xi}^k)$ and as before
$\Del T = T/(N+1)$.

Representing once more the delta-functions by their Fourier decomposition,
and thereby including integrations over lagrange multipliers $(\lb_k,s_k)$ in
the measure $\prod_k dZ_k$, we can use eq.(\ref{10.7}) to evaluate the product
of $K_{\ve}(Z_{s+1},Z_s|\Delta T)$-factors on the right:

\be
\ba{l}
\dsp{ \prod_{s=0}^N\, K_{\ve}(Z_{s+1}, Z_s | \Del T)\, =\,  \left[
\frac{1}{i}\, \lh \frac{m}{4\pi^2 e_0 \Del T} \rh^2 \right]^{N+1}\, \times }
   \\
   \\
\dsp{ \times
  e^{ i \sum_{s=0}^N \Del T \left[ \frac{m}{2 e_s}
  \lh \frac{x_{s+1} - x_s}{\Del T} \rh^2 - \frac{e_s}{2} \lh m - i\ve \rh +
  i \frac{m}{e_s} \chi_{s+1} \psi_s \cdot \lh \frac{x_{s+1} - x_s}{\Del T} \rh
  + i m \chi_{s+1} \eta_s + \lb_{s+1} \lh \frac{e_{s+1} - e_s}{\Del T} \rh
  + i s_{s+1} \lh \frac{\chi_{s+1} - \chi_s}{\Del T} \rh \right]} } \\
  \\
\dsp{ \times
  e^{i \sum_{s=0}^N \Del T \left[
  i \lh \frac{b_{s+1} - b_s}{\Del T} \rh \lh \frac{c_{s+1} - c_s}{\Del T}\rh +
  \lh \frac{\bg_{s+1} - \bg_s}{ \Del T}\rh \lh \frac{\ag_{s+1} - \ag_s}{\Del T}
  \rh + i \lh \ag_{s+1} + \ag_s \rh \lh \frac{b_{s+1} - b_{s}}{\Del T} \rh
  \chi_{s+1} \right] }
}
\ea
\label{10.9}
\ee

\nit
Here the $\psi_k^{\mu}$ are defined as in eq.(\ref{7.26}), and we take
similarly

\be
\eta_k\, =\, \frac{1}{\sqrt{2m}}\, \lh \xi^3_{k+1} + \bar{\xi}^3_k \rh,
\hspace{3em}
\eta_{1k}\, =\, - \frac{i}{\sqrt{2m}}\, \lh \xi^3_{k+1} - \bar{\xi}^3_k \rh .
\label{10.10}
\ee

\nit
Also, in the front factor on the right-hand side of (\ref{10.9}) we have
included a contribution $(e_0)^{2(N+1)}$ in the denominator, instead of
$\prod_{k=0}^N (e_k)^{2}$, by making use of the fact that $e_k$ is constant:
in the integral its value remains the same from time step to time step.

Finally, the exponent of the finite differences in $(\xi_k, \bar{\xi}_k)$
in eq.(\ref{10.8}) can in the continuum limit be rewritten in terms of the
$\psi^{\mu}$, $\eta$ and $\eta_1$ and their derivatives:

\be
\sum_{j = 1}^N\, \left[ \lh \bar{\xi}_j - \bar{\xi}_{j-1} \rh \cdot \xi_j -
\bar{\xi}_j \cdot \lh \xi_{j+1} - \xi_j \rh \right]\: \rightarrow\:
 - m \lh \psi \cdot \dot{\psi} + \eta \dot{\eta} + \eta_1 \dot{\eta}_1 \rh.
\label{10.11}
\ee

\nit
As a result we can now construct the propagator of the theory in the
full ghost-extended state space as

\be
\Del_{gf}(Z,Z^{\prime})\, =\, \frac{ie}{2m}\, \int_0^{\infty} dT\,
 \int_{\Gam} DZ(\tau)\, e^{i S_{gf}[Z(\tau)]},
\label{10.12}
\ee

\nit
where integration over the lagrange multipliers $(\lb(\tau),s(\tau))$ is to be
included in the measure $\cD Z(\tau)$, and modulo fermionic boundary terms
the action $S_{gf}$ is that of eq.(\ref{8.14}) with the addition of a single
topological fermion of the kind (\ref{8.7.1}). As has been discussed before,
at the classical level this additional fermion is completely harmless, whilst
in the quantum theory it signifies the doubling of the number of components
of the spinor wave-functions, or equivalently the doubling of the number of
degrees of freedom in the propagator.

\section{Conclusions}{\label{S.11}}

In this paper we have studied the propagators of free spin-0 and spin-1/2
particles and connected them to classical relativistic particle-mechanics
through the path-integral formalism. It has been established that starting
from the known field-theoretical expressions is advantageous, as it can
specify the representation of certain operators to be used, something
which is usually not possible from the canonical `Poisson-bracket to
commutator' quantization prescription. In the case of Dirac fermions,
this has been used to argue in favour of the bosonic representation of the
operator $\hps_5$, implying that non-manifestly supersymmetric models are
prefered to avoid doubling of the spectrum of states. Another way to resolve
this problem would be to project out half of the states by additional
constraints. Such an approach, in which a superselection rule is invoked
to restrict the physical matrix elements to half of the spinor degrees
of freedom, has been attempted for example in \ct{bordi}.

Our results can be generalized to the case of particles moving in
certain background fields: scalar, vector (e.g., electro-magnetic)  or
gravitational fields can be included in the path-integral expressions
for the propagators. The constructions we have presented have been
designed in such a way that only minimal additional work is needed
to cover these more general cases. An example is the inclusion of scalar
fields, which by Yukawa couplings may give rise to mass generation through
the mechanism of spontaneous symmetry breaking. The contribution of the
scalar interactions to the classical action (\ref{8.6}) is obtained by
replacing the terms for the fermionic multiplet $(\eta, f)$, which for
free particles read $i \eta \dot{\eta}  + 2 i \chi \eta + e f^2 - 2 e f$, by

\be
\Del L_{sc}\, =\, i \eta \dot{\eta}\, +\, e f^2\, -\, 2 g e f \vf(x)\, -\,
 2 i g e \eta \psi^{\mu} \pl_{\mu} \vf(x)\, +\, 2 i g \chi \eta \vf(x).
\label{11.1}
\ee

\nit
Here $\vf(x)$ is the scalar field, and $g$ is the Yukawa coupling constant.
Eliminating the auxiliary variable $f$ by splitting off a square, this becomes

\be
\Del L_{sc}\, =\, i \eta \dot{\eta}\, -\, eg^2 \vf^2\, -\,
 2 i g e \eta \psi^{\mu} \pl_{\mu} \vf\, +\, 2 i g \chi \eta \vf.
\label{11.2}
\ee

\nit
This result, here obtained through multiplet calculus, agrees with the result
derived by dimensional reduction in \ct{mondr}. Taking $\vf(x) = const.\ \neq
0$ returns us to the original action for a free massive particle, and shows
how masses are generated by spontaneous symmetry breaking. It is quite
straightforward to apply these results to the other actions for scalar
and Dirac particles, by either removing all fermionic degrees of freedom,
or by keeping them whilst replacing the fermionic $\eta$ by the bosonized
$\psi_5$.

Interactions with the electro-magnetic or other vector fields can be
introduced, e.g.\ through minimal coupling, whilst gravitational interactions
result from covariantizing the expressions with respect to general co-ordinate
transformations. A quantum treatment of the point particle in curved space has
been presented in \ct{vn1}. A general discussion of the inclusion of
background fields will be presented in a separate paper \ct{jw8}.
\vs

\nit
{\bf Acknowledgement}
\vs

\nit
The research described in this paper is supported in part by the
Human Capital and Mobility program of the European Union through the
network on Constrained Dynamical Systems.

\end{document}